\documentclass[twocolumn,numberedappendix,iop]{openjournal}

\usepackage{lmodern, microtype}
\usepackage{graphicx,amsmath,amssymb,amstext,stmaryrd}
\usepackage{amsbsy,amsfonts,amsthm,color}
\usepackage[colorlinks,linkcolor=blue,citecolor=blue,urlcolor=blue]{hyperref}
\usepackage[utf8]{inputenc}
\usepackage{xcolor}
\usepackage[T1]{fontenc}
\usepackage{fontawesome5}
\usepackage[title]{appendix}
\usepackage{tensor}
\usepackage{ifthen}
\usepackage{import}
\usepackage{orcidlink}
\usepackage{xspace}
\usepackage{cleveref}
\crefname{section}{sec.}{secs.}
\Crefname{section}{Section}{Sections}
\crefname{subsection}{subsec.}{subsecs.}
\Crefname{subsection}{Subsection}{Subsections}
\crefname{subsubsection}{subsubsec.}{subsubsecs.}
\Crefname{subsubsection}{Subsubsection}{Subsubsections}
\newcommand\ph{\ensuremath{\varphi}}
\newcommand\eps{\ensuremath{\varepsilon}}

\newcommand\define{\equiv}

\newcommand\vect[1]{\boldsymbol{#1}}
\newcommand\uvect[1]{\hat{\vect{#1}}}

\newcommand\ex[1]{\mathrm{e}^{#1}}
\newcommand\ii{\mathrm{i}}

\newcommand\e[1]{_{\mathrm{#1}}}

\newcommand{\dd}{\mathrm{d}}

\newcommand{\delimiters}[4][]{
\ifthenelse{ \equal{#1}{1} }{  #2 #3 #4  }
					{ \ifthenelse{\equal{#1}{2}}{ \big#2 #3 \big#4 }
						{ \ifthenelse{\equal{#1}{3}}{ \Big#2 #3 \Big#4 }
							{ \ifthenelse{\equal{#1}{4}}{ \bigg#2 #3 \bigg#4 }
								{ \ifthenelse{\equal{#1}{5}}{ \Bigg#2 #3 \Bigg#4 }
									{ \left#2 #3 \right#4 }
								}
							}
						}
					}
													}
													
\newcommand{\pa}[2][]{\delimiters[#1]{(}{#2}{)}}
\newcommand{\pac}[2][]{\delimiters[#1]{[}{#2}{]}}
\newcommand{\paac}[2][]{\delimiters[#1]{\{}{#2}{\}}}

\definecolor{blue4}{RGB}{0,0,143}
\definecolor{red4}{RGB}{143,0,0}
\definecolor{orange}{RGB}{255,128,0}
\definecolor{darkcyan}{RGB}{0,128,128}
\definecolor{olive}{RGB}{0,128,0}
\definecolor{purple}{RGB}{128,0,128}
\definecolor{cyan2}{RGB}{0,255,255}
\definecolor{fushia}{RGB}{255,0,255}
\definecolor{mygray}{gray}{0.6}
\definecolor{lightgray}{gray}{0.85}

\newcommand{\pairs}{\mathcal{P}}
\newcommand{\redshiftbin}{\mathcal{R}}
\newcommand{\univ}{\mathcal{U}}
\newcommand{\Cov}{\mathrm{Cov}}
\newcommand{\CCov}{\mathrm{CCov}}
\newcommand{\SCov}{\mathrm{SCov}}
\newcommand{\NCov}{\mathrm{NCov}}

\newcommand{\cev}[2][]{\delimiters[#1]{\langle}{#2}{\rangle}}
\newcommand{\sev}[2][]{\delimiters[#1]{\llbracket}{#2}{\rrbracket}}
\newcommand{\footprint}{\mathcal{F}}
\newcommand{\annulus}{\mathcal{A}}

\newcommand{\new}[1]{\textcolor{olive}{#1}}
\newcommand{\negligible}[1]{\textcolor{mygray}{#1}}

\newcommand{\notebook}{\href{https://github.com/pierrefleury/sparsity-covariance/blob/main/sparsity_covariance.ipynb}{\faGithub}}
\newcommand{\repository}{\href{https://github.com/pierrefleury/sparsity-covariance}{\faGithub}}
\newcommand{\OneCov}{\textsc{OneCovariance}}

\newcommand{\correction}[1]{#1}

\begin{document}

\title{Sparsity covariance: a source of uncertainty when estimating correlation functions with a discrete sample of observations in the sky
\vspace{-4em}}
\shorttitle{Sparsity covariance}

\author{Pierre Fleury \orcidlink{0000-0001-9292-3651}}
\shortauthors{Pierre Fleury}

\thanks{E-mail: \href{mailto:pierre.fleury@lupm.in2p3.fr}{pierre.fleury@lupm.in2p3.fr}}
\affiliation{
Laboratoire Univers et Particules de Montpellier (LUPM),
CNRS \& Université de Montpellier (UMR-5299),\\
Parvis Alexander Grothendieck, F-34095 Montpellier Cedex 05, France
}

\begin{abstract}
Cosmological observables rely heavily on summary statistics such as two-point correlation functions. In many practical cases (e.g. the weak-lensing cosmic shear), those correlation functions are estimated from a finite, discrete sample of measurements that are randomly distributed in the sky. The result then inevitably depends on the sample at hand, regardless of any experimental noise.
This sample dependence is a source of uncertainty for cosmological observables which I call sparsity covariance.
This article proposes a mathematical definition and a generic method to compute sparsity covariance.
It is then applied to the concrete case of cosmic shear, showing that sparsity covariance mostly enhances shape noise, whose amplitude is determined by the apparent ellipticity of galaxies rather than their intrinsic ellipticity. In general, sparsity covariance is non-negligible when the signal-to-noise ratio of individual measurements in the sample is comparable to, or larger than, unity.
\\[1em]
\textit{Keywords: cosmology; statistics; covariance matrix; large-scale structure; cosmic shear} 
\end{abstract}

\maketitle

\section{Introduction}

According to the current understanding of cosmology, inhomogeneities and their structure originated from quantum fluctuations in the early Universe~\citep{Peter:2013avv}. In this picture, physical fields may be seen as realisations of some Gaussian random fields that have subsequently evolved according to the laws of physics.

In order to extract the properties of those primordial random fields, it is customary to rely on summary statistics, such as the two-point correlation of astronomical measurements. In many practical cases, the correlation function is estimated from a discrete sample of measurements corresponding to observations made in more or less random directions in the sky. Examples include the weak-lensing cosmic shear~\citep{1993ApJ...404..441K, KiDS:2020suj, DES:2021vln, Dalal:2023olq}, based on the two-point correlation of the apparent ellipticity of galaxies; baryon-acoustic-oscillation measurements exploiting the two-point correlation of the Lyman-$\alpha$ absorption in quasar spectra~\citep{BOSS:2014hwf,DESI:2024lzq}; or measurements of the stochastic gravitational-wave background with pulsar timing arrays (PTA)~\citep{1983ApJ...265L..39H, NANOGrav:2023gor, Reardon:2023gzh, EPTA:2023sfo, Xu:2023wog}.

Such estimates of summary statistics inevitably depend on the samples at hand, and more specifically on their distribution in the sky. This dependence is a source of uncertainty that is independent of experimental noise. In the context of cosmic shear, for example, even if galaxies were intrinsically spherical, so that we could perfectly measure the effect of weak lensing on their apparent shape, the measured two-point correlation function would still depend on their exact distribution in the sky, resulting in an uncertainty that is distinct from shape noise, cosmic variance or even source-clustering bias~\citep{Bernardeau:1997tj}.

This statistical source of uncertainty is what I shall refer to as \emph{sparsity covariance} from now on. The present article aims to precisely define sparsity covariance, establish a general mathematical framework to compute it, and apply it to the concrete case of cosmic shear for illustration.

To the best of my knowledge, sparsity covariance has never been studied in the literature. However, the general question of how to estimate summary statistics with discrete data sets has received some attention. In the context of galaxy surveys, \cite{BaleatoLizancos:2023jbr, Euclid:2024xqh, Wolz:2024dro} developed methods to estimate angular power spectra directly from discrete observations, \correction{and \cite{Hall:2025jxt} considered the effect of pixelisating such discrete data;} the bias of the underlying estimators was computed, but not their covariance. In the context of PTA, \cite{Pitrou:2024scp} analysed the bias on the angular spectrum of the Hellings--Downs correlation due to the finite number of observed pulsars; in this case, however, the issue lies specifically in the transformation from real space to harmonic space -- if I measure the correlation function of a random field at some specific angular separations, how well can I reconstruct its multipoles?
This is quite different from sparsity covariance.\footnote{
In fact, and perhaps counter-intuitively, there is no sparsity covariance for PTAs in the limit of long pulsar monitoring times. In other words, in the absence of noise, the Hellings--Downs correlation is measured exactly for each pair of pulsars, because temporal averaging plays the role of cosmic averaging. This may seem at odds with the notion of pulsar variance introduced by \cite{Allen:2022dzg}, but the latter appears for a finite number of gravitational-wave sources, which makes the signal statistically anisotropic.}

The article is organised as follows. In sec.~\ref{sec:sparsity_covariance}, I define the notion of sparsity covariance both intuitively and mathematically, and I explain how to compute it in the general case of a Gaussian random field on the two-sphere. I then apply these concepts and methods to the case of cosmic shear in sec.~\ref{sec:application_cosmic_shear}. I summarise and conclude in sec.~\ref{sec:conclusion}. I also propose short answers to frequently asked questions~(FAQ) hereafter.

The data files and the Jupyter notebook used to produce the figures in this article are available on GitHub \repository.

\subsection*{FAQ}

\begin{itemize}
\item \textbf{What is sparsity covariance?} When estimating the statistical properties of a field that is measured on a finite, discrete sample, the result necessarily depends on this very sample. Sparsity covariance quantifies the uncertainty specifically associated with the \emph{discreteness} of the sample. This article focuses on sparsity covariance in the estimation of two-point correlation functions.
\item \textbf{How is it different to cosmic covariance and sample covariance?} Cosmic or sample covariance refer to the uncertainty in any estimate of the statistical property of a physical field due to the finite \emph{extension} of the observable universe, or of the survey. Sparsity covariance thus adds to it, due to the fact that such a finite region of the Universe is randomly sampled. This distinction is made explicit in  \cref{eq:cosmic_sparsity_covariances}, and illustrated in \cref{fig:illustration_sparsity_covariance}.
\item \textbf{How is it different to shot noise?} Shot noise emerges when measuring a physical quantity by counting things (electrons, photons, galaxies, etc.); the result is then subject to Poisson fluctuations. Sparsity covariance applies to quantities that may be intrinsically continuous, but that are \emph{evaluated} on a discrete sample. This distinction is, arguably, a mere matter of terminology.
\item \textbf{When should I care?} Sparsity variance is typically $\sim \xi^2(0)/P$, where $\xi$ is the field's correlation function and $P$ the number of pairs of measurements used to estimate it. It is non-negligible compared to cosmic variance if $\xi^2(0)/P$ is comparable or larger than the celestial average of $\xi^2(\theta)$. In the presence of noise, sparsity covariance matters if the signal-to-noise ratio of the individual measurements constituting the sample is comparable to unity, \correction{irrespective of the number of measurements.}
\item \textbf{Does it matter for cosmic shear?} Since the weak-lensing cosmic shear is estimated on a discrete sample of galaxy observations, it is, in principle, subject to sparsity covariance. \correction{In sec.~\ref{sec:application_cosmic_shear}, I show that sparsity covariance acts as an extra contribution to shape noise, implying that its net amplitude should be determined by the \emph{apparent}, rather than intrinsic, ellipticity of galaxies. This is already the case in practice, because the intrinsic ellipticity of galaxies is not observable; therefore, sparsity covariance is already, though accidentally, taken into account in cosmic shear analyses.}
\end{itemize}

\section{What is sparsity covariance?}
\label{sec:sparsity_covariance}

This section defines sparsity covariance, and introduces the relevant conceptual and mathematical tools to compute it in practice. I choose to focus on a specific problem, namely the estimation of the two-point correlation function of a random scalar field on the two-sphere, as this strikes a balance between mathematical simplicity and practical relevance.

\subsection{Set-up: a finite sample of observations of a field}

Consider a realisation of a random field~$f$ on the celestial sphere, for which we have a finite, discrete set of noiseless observations~$(f_1, \ldots, f_N)$ made in different directions in the sky.
We aim to estimate the two-point correlation function~$\xi$ of $f$ from this discrete set of observations, as well as the uncertainty on this estimate.

Throughout this section, I shall make the following \textbf{assumptions}:
\begin{enumerate}
\item $f$ is a real-value scalar field on the two-sphere, $f:\mathcal{S}^2\to\mathbb{R}$; as such, it only depends on the direction~$\uvect{u}\in\mathcal{S}^2$, or the corresponding polar coordinates~$(\theta, \ph)$ in which it is evaluated.
\label{assumption:S2_to_R}
\item $f$ has vanishing expectation value, $\cev{f}=0$.
\item $f$ is statistically isotropic: its correlation between two directions $\uvect{u}_1, \uvect{u}_2$ only depends on the angle $\theta_{12}$ formed by those directions, $\cev{f(\uvect{u}_1) f(\uvect{u}_2)}=\xi(\theta_{12})$.
\end{enumerate}
The above assumptions are made for the sake of simplicity, so as to efficiently convey the main point. In sec.~\ref{sec:application_cosmic_shear}, with the complete example of cosmic shear, I shall show how to relax assumption~\ref{assumption:S2_to_R} and work with a complex, spin-two field with an additional redshift dependence.

\subsection{Estimator of the correlation function}
\label{subsec:estimator_correlation_function}

In order to estimate the two-point correlation function of $f$, we may consider pairs $(\uvect{u}_i, \uvect{u}_j)$ of directions and bin them depending on their angular separation~$\theta_{ij}$. Let us take $A$ such bins $[\theta_a, \theta_{a+1})_{a=1,\ldots, A}$, and denote with
\begin{equation}
\pairs_a
\define
\left\lbrace
(ij) \; | \; \theta_{ij} \in [\theta_a, \theta_{a+1})
\right\rbrace
\end{equation}
the \emph{symmetrised} set of pairs of observations that fall into the $a$\textsuperscript{th} bin of angular separations. `Symmetrised' means here that we are double-counting each pair, in the sense that if $(ij)\in\pairs_a$, then $(ji)\in\pairs_a$ too. This choice eases the subsequent calculations. \correction{Hence, the cardinality of $\pairs_a$},
\begin{equation}
\correction{P_a \define |\pairs_a|} \ ,
\end{equation}
is \emph{twice} the number of pairs of directions in the $a$\textsuperscript{th} bin.

In such conditions, a simple estimator of the two-point correlation~$\xi$ of $f$ in the $a$\textsuperscript{th} bin is
\begin{equation}
\label{eq:estimator_xi}
\hat{\xi}_a
\define
\frac{1}{P_a} \sum_{(ij) \in\pairs_a} f_i f_j \ ,
\end{equation}
with $f_i\define f(\uvect{u}_i)$. A more general definition would involve non-trivial weights assigned to each direction, or each pair of directions.  A suitable choice of weights may lower the overall variance of the estimator~\citep{Pitrou:2024scp}, but this goes beyond the scope of the simple presentation intended here.

\subsection{General idea of sparsity covariance}
\label{subsec:general_idea_sparsity_covariance}

Sparsity covariance stems from the intuitive fact that the estimated correlation~$\hat{\xi}_a$ depends on the finite, sparse sample of directions~$(\uvect{u}_1, \ldots, \uvect{u}_N)$ in which the field is actually observed. As we only make use of a single, finite sample, this sample-dependence is thus a source of uncertainty in our estimation of the field's correlation. In the theoretical limit of an infinite sample, $N\to\infty$, sparsity covariance vanishes if the sky is evenly covered.

This notion of covariance must be distinguished from the so-called cosmic covariance, which comes from the fact that we observe a single sky, and hence a single \emph{realisation} of the random field~$f$. In particular, cosmic covariance would persist even if we had an infinite sample.

\begin{figure*}
\centering
\includegraphics[width=\textwidth]{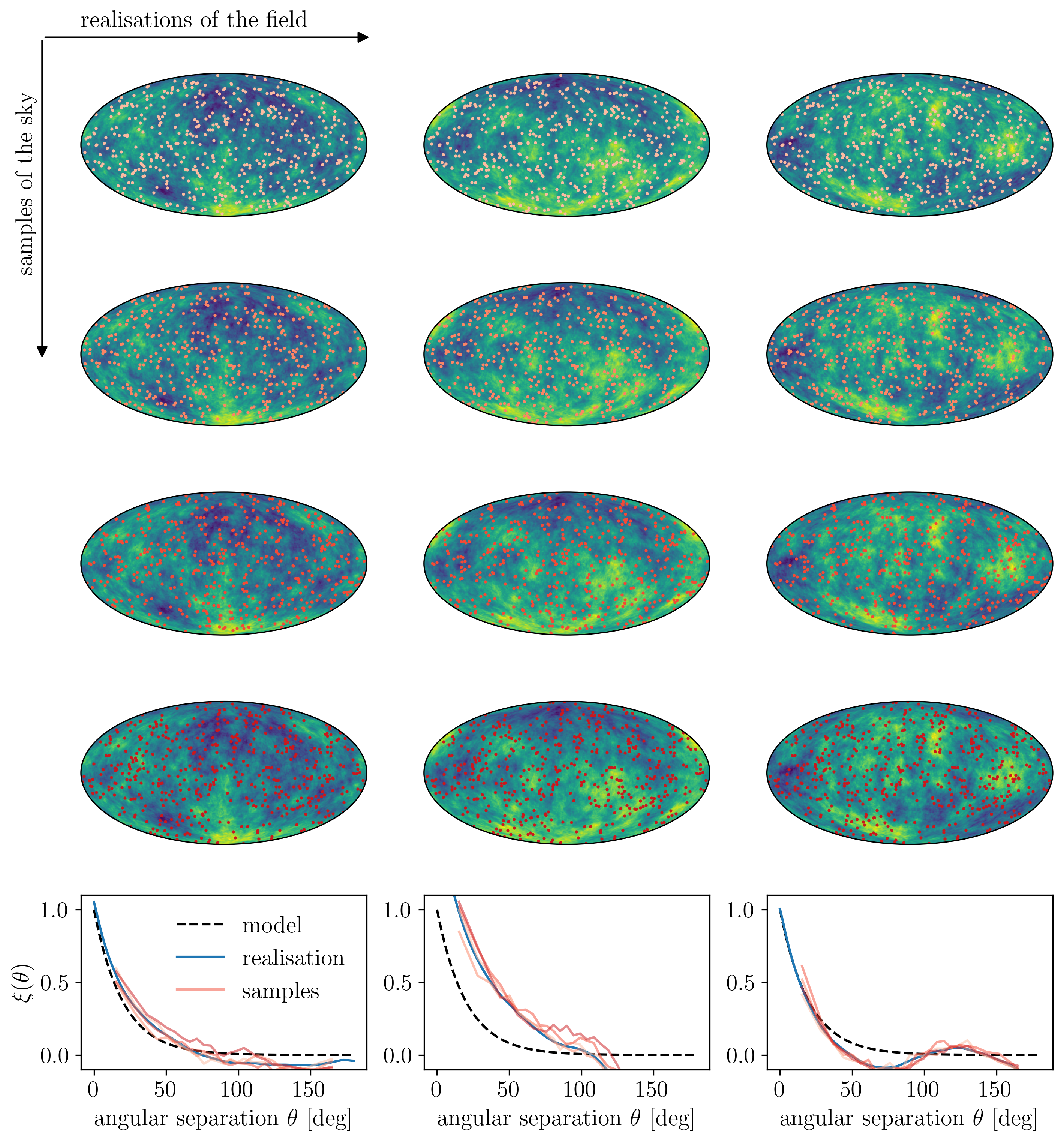}
\caption{Illustrating sparsity and cosmic covariances with an example. In the first four rows from the top, each panel is a Mollweide map of a realisation of the field $f$ (blue to yellow colours), together with a discrete sample of 500 random directions in the sky (orange to red dots). More specifically, each column corresponds to one realisation of the field, while each row corresponds to one sample of the sky. The bottom row of the figure compares various correlation functions: in each panel, the black dashed lines indicates the fundamental~$\xi(\theta)$ of the Gaussian model from which field realisations are drawn; the blue solid line is the exact correlation function of the column's realisation; and the semi-transparent orange to red lines are the estimates~$\hat{\xi}_a$ from each sample. 
Sparsity variance is the jitter of the semi-transparent red lines around the blue lines, while cosmic variance is the jitter of the blue lines around the dashed black line. \notebook}
\label{fig:illustration_sparsity_covariance}
\end{figure*}

I show an example, aimed at illustrating sparsity covariance and its difference with cosmic covariance, in \cref{fig:illustration_sparsity_covariance}. In this example, $f$ is an isotropic Gaussian random field with correlation function $\xi(\theta)=\exp(-\theta/\lambda)$, and $\lambda=20^\circ$. The first four rows of the figure are Mollweide projections of a realisation of the field (blue and yellow colours), superimposed with a random sample of $N=500$ directions of the sky (orange to red dots).  Each of the three columns of \cref{fig:illustration_sparsity_covariance} corresponds to a single realisation of the random field, while each of the first four rows corresponds to a single sample of directions. The bottom row depicts a number of correlation functions: in each panel, the black dashed line is the theoretical $\xi(\theta)$ from which the realisations are drawn; the blue solid line is the actual correlation function for this column's realisation of the field; the orange to red semi-transparent lines are the estimates~$\hat{\xi}_a$ obtained from each sample. In this illustration, the $N(N-1)/2=124750$ different pairs of distinct directions are divided into 25 equally populated bins (on average) covering separations from $0$ to $180^\circ$. This is ensured by $\cos\theta_{a}-\cos\theta_{a+1}=2/25$, so that each bin contains on average $P_a/2 = 4990$ pairs.

In \cref{fig:illustration_sparsity_covariance}, bottom row, sparsity variance is visible in each panel as the jitter of the red semi-transparent lines (sample-estimates of the correlation) around the blue line (exact correlation for a realisation). Cosmic covariance, on the other hand, would be the jitter of the solid blue lines around the black dashed line if we were to superimpose the panels. Both cosmic and sparsity covariance are sources of uncertainty when estimating the correlation function from a finite sample of a single realisation of the field.

\subsection{Definition and expression of sparsity covariance}

I now turn to a more technical discussion, explicitly defining sparsity covariance and deriving an analytical expression to predict its value for any given application.

\subsubsection{Cosmic and sample averaging}

Since this discussion involves two distinct notions of randomness, I must start with defining the corresponding notions of average. The first one, denoted with $\cev{\ldots}$, is the averaging over realisations of the field. Since we eventually aim at applications in cosmology, where $f$ is some cosmic field, $\cev{\ldots}$ may by extension correspond to averaging over realisations of the Universe, and hence be referred to as \emph{cosmic averaging}.
In practice, cosmic averaging will only be used to define correlation functions, such as in $\xi(\theta_{12})=\cev{f(\uvect{u}_1) f(\uvect{u}_2)}$.

If we are interested in a quantity~$X$ that explicitly depends on observations made in a discrete set of directions $(\uvect{u}_1, \ldots, \uvect{u}_n)$ in the sky -- such as the estimator of the correlation function defined in \cref{eq:estimator_xi} -- then \emph{sample averaging} corresponds to an average over all such sets of directions,
\begin{equation}
\sev{X}
\define
\int \dd^2\uvect{u}_1
\ldots
\dd^2\uvect{u}_n \; p(\uvect{u}_1, \ldots \uvect{u}_n|\univ) \, X(\uvect{u}_1, \ldots, \uvect{u}_n) \ ,
\end{equation}
where $p(\uvect{u}_1, \ldots \uvect{u}_n|\univ)$ denotes the probability density of the directions in the sample, given a realisation~$\univ$ of the Universe. We indeed expect sampling to generally depend on some cosmic field. For example, if our observable~$X$ is related to the Lyman-$\alpha$ absorption in quasars, then $(\uvect{u}_1, \ldots \uvect{u}_n)$ stands for the set of quasar lines of sight, which are more likely to be observed in regions of the sky where those sources are more abundant, which in turn correlates with the overall density field in $\univ$. Another realisation of the Universe would therefore lead to a different probability distribution for those lines of sights.

Having said that, however, I shall make another simplifying \textbf{assumption} in the remainder of this article. Namely, I assume that the observation directions are independent and randomly distributed with a homogeneous distribution across the footprint~$\footprint$ of some survey,
\begin{equation}
p(\uvect{u}_1, \ldots \uvect{u}_n|\univ) = \prod_{i=1}^N p(\uvect{u}_i) \ ,
\quad
p(\uvect{u})
=
\begin{cases}
\Omega^{-1} & \uvect{u}\in\footprint,\\
0 & \text{otherwise},
\end{cases}
\end{equation}
where $\Omega$ is the solid angle covered by $\footprint$, i.e. the survey area. Doing so, I choose to neglect what is usually referred to as clustering bias in the cosmology literature -- see e.g. \cite{Bernardeau:1997tj} in the context of cosmic shear. As we shall see in the following, this simplified framework is sufficient to capture the essence of sparsity covariance while making practical calculations much easier to handle.

A notable advantage of the assumption of statistically homogeneous sampling is that, since the underlying distribution does not depend on the cosmic realisation~$\univ$ any longer, the two averaging procedures commute,
\begin{equation}
\sev{\cev{X}} = \cev{\sev{X}} .
\end{equation}
%

\subsubsection{Defining cosmic and sparsity covariances}

Equipped with the distinct notions of cosmic and sample averaging, we are ready to define the cosmic and sparsity covariances of the estimator~\eqref{eq:estimator_xi} of the correlation function. The full covariance matrix is defined as the average of the product minus the product of averages; for any two angular bins $a, b$,
\begin{equation}
\label{eq:covariance_definition}
\Cov(\hat{\xi}_a, \hat{\xi}_b)
\equiv \cev[2]{\sev[2]{\hat{\xi}_a\hat{\xi}_b}} - \cev[2]{\sev[2]{\hat{\xi}_a}} \cev[2]{\sev[2]{\hat{\xi}_b}} .
\end{equation}
Adding and subtracting the term~$\cev[2]{\sev[2]{\hat{\xi}_a} \sev[2]{\hat{\xi}_b}}$, we may then split the above into two contributions,
\begin{align}
\Cov(\hat{\xi}_a, \hat{\xi}_b)
&=
\underbrace{
\cev[2]{\sev[2]{\hat{\xi}_a\hat{\xi}_b} - \sev[2]{\hat{\xi}_a} \sev[2]{\hat{\xi}_b}}
}_{\text{sparsity covariance}~\SCov(\hat{\xi}_a, \hat{\xi}_b)}
\nonumber \\
&\quad +
\underbrace{
\cev[2]{\sev[2]{\hat{\xi}_a} \sev[2]{\hat{\xi}_b}}
- \cev[2]{\sev[2]{\hat{\xi}_a}} \cev[2]{\sev[2]{\hat{\xi}_b}}
}_{\text{cosmic covariance}~\CCov(\hat{\xi}_a, \hat{\xi}_b)} \ ,
\label{eq:cosmic_sparsity_covariances}
\end{align}
which defines the two types of covariance.

Let me take a moment to connect those expressions to the illustration of \cref{fig:illustration_sparsity_covariance}. For a given realisation of the field, $\sev[1]{\hat{\xi}_a}$ is nothing but the binned correlation function of that realisation; it therefore corresponds to the blue solid line in each bottom panel of \cref{fig:illustration_sparsity_covariance}. It follows that $\sev[1]{\hat{\xi}_a^2} - \sev[1]{\hat{\xi}_a}^2$ is the variance of the fluctuations of the red semi-transparent lines around each blue line, which is sparsity variance. To be really precise, $\cev[1]{\sev[1]{\hat{\xi}_a^2} - \sev[1]{\hat{\xi}_a}^2}$ is the cosmic average of sparsity variance, that is the average jitter across realisations of the field. Cosmic covariance is easier to identify: since $\sev[1]{\hat{\xi}_a}$ corresponds to the blue curves, $\cev[1]{\sev[1]{\hat{\xi}_a}^2} - \cev[1]{\sev[1]{\hat{\xi}_a}}^2$ is simply the variance of their fluctuations around the black dashed line.

Albeit theoretically eloquent, the separation of terms in \cref{eq:cosmic_sparsity_covariances} is not necessarily the most convenient one in practice. An alternative is to recall that cosmic covariance is the contribution to $\Cov(\hat{\xi}_a, \hat{\xi}_b)$ that persists when the sample size goes to infinity,
\begin{equation}
\CCov(\hat{\xi}_a, \hat{\xi}_b)
\equiv
\lim_{N\to\infty} \Cov(\hat{\xi}_a, \hat{\xi}_b) \ ,
\end{equation}
while sparsity covariance is the remainder,
\begin{equation}
\SCov(\hat{\xi}_a, \hat{\xi}_b)
\equiv
\Cov(\hat{\xi}_a, \hat{\xi}_b)  - \CCov(\hat{\xi}_a, \hat{\xi}_b) \ .
\end{equation}

In the following, I derive analytical expressions for cosmic and sampling covariances. As it turns out, the easiest path starts with using the (\correction{generally} approximate) commutation of cosmic and sample averaging, in order to rewrite the full covariance of \cref{eq:covariance_definition} as
\begin{equation}
\label{eq:covariance_convenient_order}
\Cov(\hat{\xi}_a, \hat{\xi}_b)
=
\sev[2]{\cev[2]{\hat{\xi}_a\hat{\xi}_b}} - \sev[2]{\cev[2]{\hat{\xi}_a}} \sev[2]{\cev[2]{\hat{\xi}_b}} .
\end{equation}
The advantage of this expression is that the cosmic average is applied before the sample average, which henceforth allows me to work with deterministic functions instead of random fields. 

\subsubsection{Expectation value of the estimator}
\label{subsubsec:expectation_value_estimator}

As an illustration of the above point, let me first consider the expectation value of $\hat{\xi}$. Its cosmic average reads
\begin{equation}
\label{eq:cosmic_average_estimator}
\cev[2]{\hat{\xi}_a}
=
\frac{1}{P_a} \sum_{(ij) \in\pairs_a} \xi_{ij} \ ,
\end{equation}
with
\begin{equation}
\xi_{ij}
=
\xi(\theta_{ij})
\define
\cev{f_i f_j} .
\end{equation}
When averaging \cref{eq:cosmic_average_estimator} over samples, the number of pairs~$P_a$ is in principle subject to Poisson-like fluctuations. In the limit of large numbers, however, such fluctuations may be neglected and $P_a$ may be identified with its mean value. Besides, all the pairs~$(ij)$ are statistically equivalent. In other words, $i, j$ are dummy indices that may be arbitrarily replaced with $1, 2$, so that the right-hand side reduces to a single term,
\begin{equation}
\label{eq:sample_average_xi_one_term}
\sev[2]{\cev[2]{\hat{\xi}_a}}
=
\xi_{\sev{12}_a}
\ ,
\end{equation}
where the shorthand notation~$\xi_{\sev{12}_a}$ refers to the directional average of $\xi_{12}=\xi(\uvect{u}_1, \uvect{u}_2)$ for all possible configurations of $(\uvect{u}_1, \uvect{u}_2)$ such that $\theta_{12}$ falls in the $a$\textsuperscript{th} bin.

\begin{figure}[t]
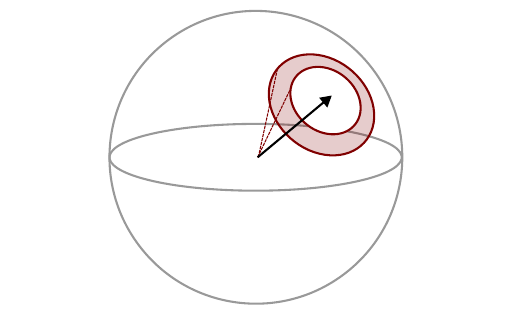
\caption{Illustration of the quantities involved in \cref{eq:expectation_value_xi_exact}. The footprint~$\footprint$ is the area of the celestial sphere covered by some survey. The annulus~$\annulus_a(\uvect{u}_1)$ materialises the $a$\textsuperscript{th} bin of angular separation, in the sense that $(12)\in\pairs_a\Leftrightarrow\theta_{12}\in[\theta_a, \theta_{a+1})\Leftrightarrow\uvect{u}_2\in\annulus_a(\uvect{u}_1)$. For small angular separations the annulus is almost always included in the footprint and we may neglect the curvature of the sky.
}
\label{fig:footprint_annulus}
\end{figure}

An explicit integration scheme then consists in taking $\uvect{u}_1$ to cover the whole $\footprint$, while $\uvect{u}_2$ spans an annulus~$\annulus_a(\uvect{u}_1)$ drawn around $\uvect{u}_1$ such that~$\theta_a\leq\theta_{12}<\theta_{a+1}$, provided this region remains within $\footprint$ (see \cref{fig:footprint_annulus}). If I denote with $\Omega_a(\uvect{u}_1)$ the solid angle covered by $\annulus_a(\uvect{u}_1)\cap\footprint$, then
\begin{equation}
\label{eq:expectation_value_xi_exact}
\xi_{\sev{12}_a}
=
\int_\footprint \frac{\dd^2 \uvect{u}_1}{\Omega}
\int_{\annulus_a(\uvect{u}_1)\cap\footprint} \frac{\dd^2\uvect{u}_2}{\Omega_a(\uvect{u}_1)} \;
\xi(\theta_{12})
\ .
\end{equation}
For a full-sky survey ($\footprint=\mathcal{S}^2$), $\annulus_a$ is always contained in $\footprint$, so that the $\uvect{u}_1$ dependence drops; I may then place $\uvect{u}_1$ at the North pole~$\uvect{z}$ to get the simpler expression
\begin{equation}
\xi_{\sev{12}_a}
=
\int_{\cos\theta_{a+1}}^{\cos\theta_{a}} \frac{\dd\cos\theta}{\cos\theta_{a} - \cos\theta_{a+1}} \, \xi(\theta)
\ ,
\end{equation}
since $\Omega_a=2\pi(\cos\theta_{a} - \cos\theta_{a+1})$. More generally, if one focuses on small angular separations compared to the survey footprint, the annulus only gets truncated in the immediate vicinity of $\partial\footprint$, and one may simply neglect such boundary effects. This may be referred to as the \textbf{pseudo-full-sky approximation}. Furthermore, for small angular separations~$\theta_a\ll 1$, one may also neglect the curvature of the sky so that
\begin{equation}
\xi_{\sev{12}_a}
\approx
\frac{1}{\theta_{a+1}-\theta_a} \int_{\theta_a}^{\theta_{a+1}} \dd\theta \;  \theta \, \xi(\theta)
\ ,
\end{equation}
which is the \textbf{flat-sky approximation}.

\subsubsection{The mathematical origin of sparsity covariance}
\label{subsubsec:mathematical_origin_sparsity_covariance}

Let me now tackle the quadratic term of \cref{eq:covariance_convenient_order}; starting with cosmic averaging, I have
\begin{equation}
\cev[2]{\hat{\xi}_a\hat{\xi}_b}
=
\frac{1}{P_a P_b} \sum_{(ij)\in\pairs_a} \sum_{(kl)\in\pairs_b} \Xi_{ijkl}
\end{equation}
where
\begin{equation}
\Xi_{ijkl}
=
\Xi(\uvect{u}_i, \uvect{u}_j, \uvect{u}_k,\uvect{u}_l)
\define
\cev{f_i f_j f_k f_l}
\end{equation}
denotes the four-point correlation function of $f$, which depends on the shape and size of the quadrilateral formed by the four directions $\uvect{u}_i, \uvect{u}_j, \uvect{u}_k,\uvect{u}_l$.

%

\correction{
Applying the sample-averaging operator, where I neglect again the Poisson fluctuations of the number of pairs~$P_a, P_b$, I am left with
\begin{equation}
\label{eq:sample_average_quadratic_start}
\sev[2]{\cev[2]{\hat{\xi}_a \hat{\xi}_b}}
=
\frac{1}{P_a P_b}
\sev{
\sum_{(ij) \in \pairs_a}  \sum_{(kl) \in \pairs_b} \Xi_{ijkl}
} .
\end{equation}
At this point, it is crucial to notice that,} contrary to \cref{eq:cosmic_average_estimator}, \emph{not all the terms of the double sum are equivalent}.
This is because the pairs $(ij)$ and $(kl)$ are drawn from the same sample of directions~$(\uvect{u}_1, \ldots, \uvect{u}_N)$, and hence they sometimes have one direction in common  -- sometimes even two if $a=b$.

\begin{figure}[t]
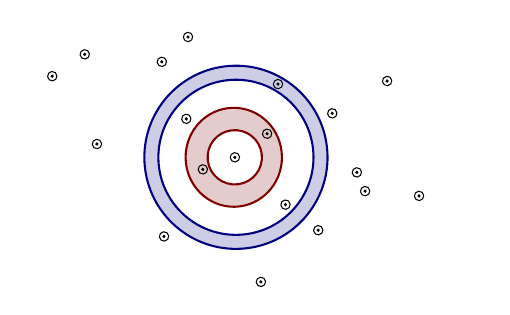
\caption{\Cref{eq:sample_average_quadratic_start} features a double sum over pairs of directions, $(ij)\in\pairs_a$ and $(kl)\in\pairs_b$. In that sum, some terms may have a direction in common, such as $(12)\in\pairs_a$ and $(14)\in\pairs_b$ which have $\uvect{u}_1$ in common. This specific configuration is represented by four terms in the double sum: $(12)(14)$; $(12)(41)$; $(21)(14)$; $(21)(41)$.}
\label{fig:common_direction}
\end{figure}

For instance, in the sample depicted in \cref{fig:common_direction}, the pair of directions~$(\uvect{u}_1, \uvect{u}_2)$ falls in the $a$\textsuperscript{th} bin, while $(\uvect{u}_1, \uvect{u}_4)$ falls in the $b$\textsuperscript{th} bin, so $(12)\in\pairs_a$ and $(14)\in\pairs_b$. Thus, in this specific configuration, the double sum of \cref{eq:sample_average_quadratic_start} features the term~$\Xi_{1214}$.  Averaging over all $\uvect{u}_1, \uvect{u}_2, \uvect{u}_4$ that respect this configuration, yields
\begin{multline}
\label{eq:Xi_1214}
\Xi_{\sev{12}_a\sev{14}_b}
=
\int_\footprint \frac{\dd^2 \uvect{u}_1}{\Omega}
\int_{\annulus_a(\uvect{u}_1)\cap\footprint} \frac{\dd^2\uvect{u}_2}{\Omega_a(\uvect{u}_1)}
\\
\int_{\annulus_b(\uvect{u}_1)\cap\footprint} \frac{\dd^2\uvect{u}_4}{\Omega_b(\uvect{u}_1)} \;
\Xi(\uvect{u}_1, \uvect{u}_2, \uvect{u}_1,\uvect{u}_4)  \ ,
\end{multline}
which is manifestly different from \correction{the case where, e.g., the four indices of $\Xi_{ijkl}$ are all distinct:}
\begin{multline}
\label{eq:Xi_1234}
\Xi_{\sev{12}_a\sev{34}_b}
=
\int_\footprint \frac{\dd^2 \uvect{u}_1}{\Omega}
\int_{\annulus_a(\uvect{u}_1)\cap\footprint} \frac{\dd^2\uvect{u}_2}{\Omega_a(\uvect{u}_1)}
\\
\int_\footprint \frac{\dd^2 \uvect{u}_3}{\Omega}
\int_{\annulus_b(\uvect{u}_3)\cap\footprint} \frac{\dd^2\uvect{u}_4}{\Omega_b(\uvect{u}_3)} \;
\Xi(\uvect{u}_1, \uvect{u}_2, \uvect{u}_3,\uvect{u}_4) \ .
\end{multline}

More generally, the $P_a P_b$ terms in the double sum of \cref{eq:sample_average_quadratic_start} may be divided into three classes:
\begin{itemize}
\item $C_0$ terms for which $i, j, k, l$ are all different (zero directions in common);
\item $C_1$ terms for which two and only two indices in $i, j, k, l$ are equal (one direction in common);
\item $C_2$ terms for which  $\{i, j\}=\{k, l\}$ (two directions in common), which of course is only possible if $a=b$; due to the double counting of pairs, this covers both $(ij)=(kl)$ and $(ij)=(lk)$.
\end{itemize}
One must have $C_0+C_1+C_2=P_a P_b$ to recover the total number of terms. Within each class, all the terms are equivalent up to a relabelling of the directions. They are, respectively, equal to $\Xi_{\sev{12}_a\sev{34}_b}$ [\cref{eq:Xi_1234}] for the terms with no index in common; $\Xi_{\sev{12}_a\sev{14}_b}$ [\cref{eq:Xi_1214}] for those with one index in common; and
\begin{multline}
\label{eq:Xi_1212}
\Xi_{\sev{12}_a \sev{12}_a}
\define
\int_\footprint \frac{\dd^2 \uvect{u}_1}{\Omega}
\int_{\annulus_a(\uvect{u}_1)\cap\footprint} \frac{\dd^2\uvect{u}_2}{\Omega_a(\uvect{u}_1)}\\
\times \Xi(\uvect{u}_1, \uvect{u}_2, \uvect{u}_1,\uvect{u}_2) 
\end{multline}
for the terms with two indices in common.

Following that division, and neglecting the Poisson fluctuations of $C_0, C_1, C_2$, \cref{eq:sample_average_quadratic_start} becomes
\begin{equation}
\sev[2]{\cev[2]{\hat{\xi}_a \hat{\xi}_b}}
=
F_0 \, \Xi_{\sev{12}_a\sev{34}_b}
+ F_1 \, \Xi_{\sev{12}_a\sev{14}_b}
+ F_2 \, \Xi_{\sev{12}_a\sev{12}_b} \ ,
\label{eq:quadratic_term_fractions}
\end{equation}
where $F_p\define C_p/P_a P_b$ denotes the fraction of terms in double sum that have $p$ indices in common.
\correction{The terms proportional to $F_1, F_2$ are the origin of sparsity covariance.}

\subsubsection{Counting pairs of pairs with points in common}
\label{subsubsec:counting_pairs_of_pairs}

The last step of the calculation is to evaluate the fractions~$F_0, F_1, F_2$ of \cref{eq:quadratic_term_fractions}. I propose here a rather intuitive derivation based on a continuous approach, but the final result is expected to be accurate for $N\gg 1$.

Let me denote with $n\define N/\Omega$ the average  density of the sample. Around each direction~$\uvect{u}_i$, there are on average $n\Omega_a$ other directions~$\uvect{u}_j$ such that $\uvect{u}_j\in\annulus_a(\uvect{u}_i)$. So the average number of pairs in $\pairs_a$ is simply
\begin{equation}
P_a = N \times n\Omega_a \ .
\end{equation}
This expression does double-count the number of actual pairs, because $(ij)$ is counted when $\uvect{u}_j\in\annulus_a(\uvect{u}_i)$ and $(ji)$ when $\uvect{u}_i\in\annulus_a(\uvect{u}_j)$.

A similar logic may be employed to determine the average number~$C_1$ of terms of the double sum over $(ij)$, $(kl)$ that have one index in common. For each pair of pairs, there are four possibilities: $i=k$, $i=l$, $j=k$ and $j=l$; I can thus write
\begin{align}
C_1
&= \sum_{(ij)\in\pairs_a} \sum_{(kl)\in\pairs_b} \pa{\delta_{ik} + \delta_{il} + \delta_{jk} + \delta_{jl}}
\\
&= 4 \sum_{(ij)\in\pairs_a} \sum_{(kl)\in\pairs_b} \delta_{ik}
\label{eq:calculation_C_1_symmetry}
\\
&= 4 \sum_{i=1}^N
\sum_{\substack{j\\(ij)\in\pairs_a}}
\sum_{\substack{l\\(il)\in\pairs_b}} 1
\end{align}
where in \cref{eq:calculation_C_1_symmetry} I used the symmetry under $i\leftrightarrow j$ and $k\leftrightarrow l$ due to double counting. On average, the sums over $j$ and $l$ in the above are respectively $n\Omega_a$ and $n\Omega_b$, so that $C_1=4N (n\Omega_a)(n\Omega_b)$, and hence
\begin{equation}
F_1 = \frac{C_1}{P_a P_b} = \frac{4}{N} \ .
\end{equation}

As for the terms with two indices in common, the calculation is even simpler. When $a\neq b$, $C_2=0$ by definition, and when $a=b$ there are two possibilities: $(ij)=(kl)$ or $(ij)=(lk)$; so
\begin{align}
C_2
&= \delta_{ab} \sum_{(ij)\in\pairs_a} \sum_{(kl)\in\pairs_a} \pa{\delta_{ik}\delta_{jl} + \delta_{il}\delta_{jk} }
\\
&= 2\delta_{ab} \sum_{(ij)\in\pairs_a} \sum_{(kl)\in\pairs_a} \delta_{ik}\delta_{jl}
\\
&= 2\delta_{ab} \sum_{(ij)\in\pairs_a} 1 = 2\delta_{ab} P_a \ ,
\end{align}
whence
\begin{equation}
F_2
= \frac{C_2}{P_a P_b}
= \frac{2 \delta_{ab}}{P_a} = \frac{2\delta_{ab}}{N^2} \frac{\Omega}{\Omega_a} \ .
\end{equation}

Substituting $F_1, F_2$, and $F_0=1-F_1-F_2$ in \cref{eq:quadratic_term_fractions}, subtracting the product of expectation values $\sev[1]{\cev[1]{\hat{\xi}_a}}\sev[1]{\cev[1]{\hat{\xi}_b}}$, and identifying cosmic covariance as the limit of the full covariance for $N\to\infty$, I finally obtain the almost explicit expressions
\begin{align}
\label{eq:cosmic_covariance_semi_explicit}
\CCov(\hat{\xi}_a, \hat{\xi}_b)
&=
\Xi_{\sev{12}_a\sev{34}_b} -  \xi_{\sev{12}_a} \xi_{\sev{34}_b} ,
\\
\label{eq:sparsity_covariance_semi_explicit}
\SCov(\hat{\xi}_a, \hat{\xi}_b)
&=
\frac{4}{N} \pa{\Xi_{\sev{12}_a\sev{14}_b} - \Xi_{\sev{12}_a\sev{34}_b}}
\nonumber\\ & +
\frac{2\delta_{ab}}{P_a}
\pa{\Xi_{\sev{12}_a\sev{12}_a} - \Xi_{\sev{12}_a\sev{34}_a}}
.
\end{align}

\subsubsection{Special case: Gaussian field}

One may go one step further by assuming that $f$ is a Gaussian random field, which allows one to express the four-point correlation function in terms of the two-point correlation function~\citep{isserlis_1918},
\begin{equation}
\label{eq:Isserlis}
\Xi_{ijkl}
=
\xi_{ij} \xi_{kl} + \xi_{ik} \xi_{jl} + \xi_{il} \xi_{jk} \ .
\end{equation}
The cosmic covariance contribution then reduces to
\begin{align}
\CCov(\hat{\xi}_a, \hat{\xi}_b)
&=
\xi\indices{^{\llbracket 1}_{\llbracket 3}}\xi\indices{^{2\rrbracket_a}_{4\rrbracket_b}}
+
\xi\indices{^{\llbracket 1}_{\llbracket 4}}\xi\indices{^{2\rrbracket_a}_{3\rrbracket_b}}
\\
&= 2 \xi\indices{^{\llbracket 1}_{\llbracket 3}}\xi\indices{^{2\rrbracket_a}_{4\rrbracket_b}} ,
\end{align}
where I used the altitude of indices to group them by pairs belonging to the same bin.\footnote{The notation $\xi_{\llbracket 1 \llbracket 3}\xi_{2 \rrbracket_a 4\rrbracket_b}$, for instance, would be ambiguous: it is unclear whether the pairs in bins $a, b$ are $(12), (34)$ or $(14), (23)$.} Explicitly, this reads
\begin{multline}
\xi\indices{^{\llbracket 1}_{\llbracket 3}}\xi\indices{^{2\rrbracket_a}_{4\rrbracket_b}} 
=
\int_\footprint \frac{\dd^2 \uvect{u}_1}{\Omega}
\int_{\annulus_a(\uvect{u}_1)\cap\footprint} \frac{\dd^2\uvect{u}_2}{\Omega_a(\uvect{u}_1)}
\\
\int_\footprint \frac{\dd^2 \uvect{u}_3}{\Omega}
\int_{\annulus_b(\uvect{u}_3)\cap\footprint} \frac{\dd^2\uvect{u}_4}{\Omega_b(\uvect{u}_3)} \;
\xi(\theta_{13}) \xi(\theta_{24})\ .
\end{multline}

In the pseudo-full-sky approximation, the dependence in, e.g., $\uvect{u}_1$ drops; I may then take that direction as a reference and fix it to the North pole, $\uvect{u}_1=\uvect{z}$, to get
\begin{multline}
\xi\indices{^{\llbracket 1}_{\llbracket 3}}\xi\indices{^{2\rrbracket_a}_{4\rrbracket_b}} 
\approx
\int_\footprint \frac{\dd^2 \uvect{u}_3}{\Omega} \; \xi(\theta_3)
\\
\times
\int_{\annulus_a} \frac{\dd^2\uvect{u}_2}{\Omega_a}
\int_{\annulus_b(\uvect{u}_3)} \frac{\dd^2\uvect{u}_4}{\Omega_b} \;
\xi(\theta_{24})\ ,
\end{multline}
with $\annulus_a\define\annulus_a(\uvect{z})$ and $\Omega_a, \Omega_b$ are always the same since boundary truncations are neglected.


\correction{Similarly, substituting the decomposition~\eqref{eq:Isserlis} of $\Xi_{ijkl}$ in \cref{eq:sparsity_covariance_semi_explicit},} the contribution of sparsity is found to read
\begin{align}
\label{eq:SCov_result_Gaussian}
&\SCov(\hat{\xi}_a, \hat{\xi}_b)
\nonumber\\
&=
\frac{4}{N}
\pac{
\xi\indices{^{\llbracket 1}_{\llbracket 1}}\xi\indices{^{2\rrbracket_a}_{4\rrbracket_b}}
+ \xi_{\sev{12}_a} \xi_{\sev{34}_b}
- \CCov(\hat{\xi}_a, \hat{\xi}_b)
}
\nonumber
\\
&\quad+
\frac{2\delta_{ab}}{P_a}
\pac{
\xi^2(0)
+ 2\xi^2_{\sev{12}_a}
- \pa{ \xi_{\sev{12}_a} }^2
- \CCov(\hat{\xi}_a, \hat{\xi}_a)
}
\end{align}
in the pseudo-full-sky approximation.\footnote{This approximation, together with statistical isotropy, implies some identities in $\SCov$,
namely 
$
\xi_{\sev{12}_a}\xi_{\sev{14}_b}
=
\xi_{\sev{12}_a}\xi_{\sev{34}_b}
=
\xi\indices{^{\llbracket 1}_{\llbracket 4}} \xi\indices{^{2\rrbracket_a}_{1\rrbracket_b}}
$,
and
$
\xi_{\sev{12}_a}\xi_{\sev{12}_a}
=
\xi\indices{^{\llbracket 1}_{\llbracket 2}} \xi\indices{^{2\rrbracket_a}_{1\rrbracket_b}}
=
\xi^2_{\sev{12}_a}.
$
}
All the abridged terms in the above expression have already been made more explicit, except the first one which I directly write here in the pseudo-full-sky approximation for simplicity,
\begin{align}
\xi\indices{^{\llbracket 1}_{\llbracket 1}}\xi\indices{^{2\rrbracket_a}_{4\rrbracket_b}}
\approx
\xi(0)
\int_{\annulus_a} \frac{\dd^2\uvect{u}_2}{\Omega_a} 
\int_{\annulus_b} \frac{\dd^2\uvect{u}_{4}}{\Omega_b} 
\;
\xi(\theta_{24}) \ .
\end{align}

\subsection{Validation on an example}

In order to test the expression~\eqref{eq:SCov_result_Gaussian} for sparsity covariance, I have applied it to the example considered in subsec.~\ref{subsec:general_idea_sparsity_covariance}, where $f$ is a Gaussian field with correlation function $\xi(\theta)=\exp(-\theta/\lambda)$, $\lambda=20^\circ$.

\begin{figure*}[t]
\centering
\includegraphics[width=\textwidth]{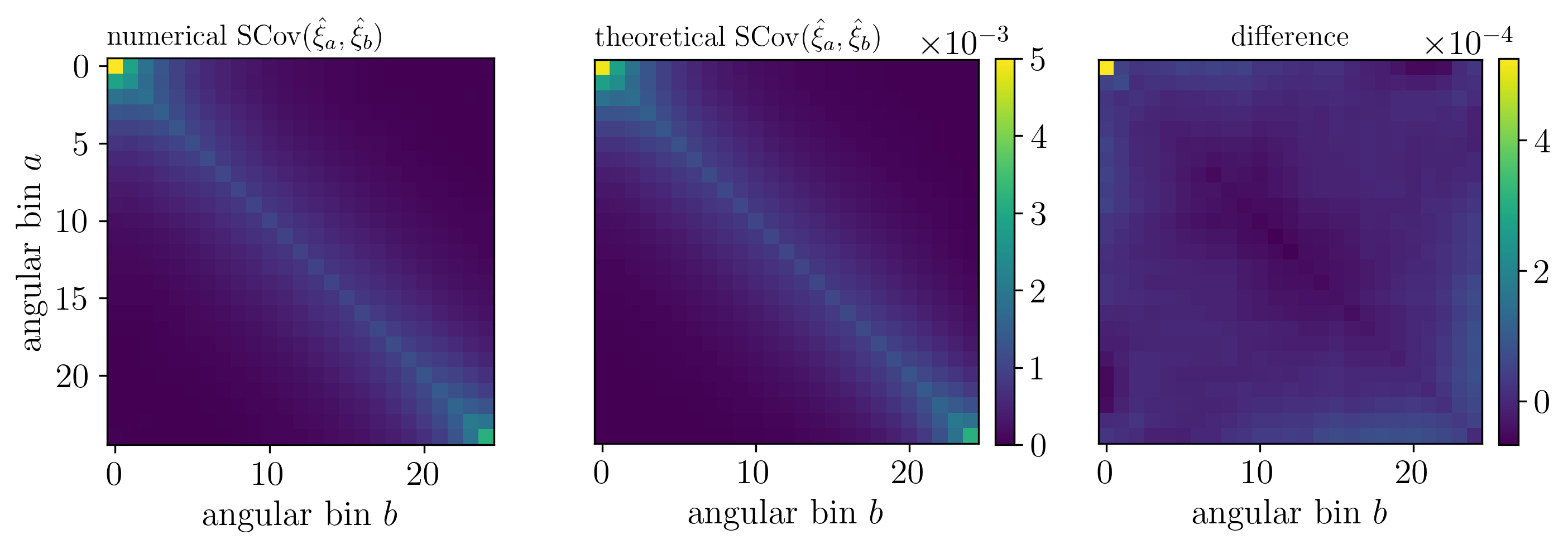}
\caption{Validation of the theoretical formula~\eqref{eq:SCov_result_Gaussian} for sparsity covariance, using the same set-up as in \cref{fig:illustration_sparsity_covariance}, i.e. a Gaussian field with correlation function~$\xi(\theta)=\exp(-\theta/20^\circ)$. The left panel is the sparsity covariance obtained from a numerical experiment, with 200 realisations of the field~$f$ and 200 random samples of 500 directions for each realisation (see main text for details); the middle panel is the theoretical prediction~\eqref{eq:SCov_result_Gaussian}; and the right panel is the difference $\text{(numerical)} - \text{(theory)}$. \notebook}
\label{fig:test_sparsity_covariance}
\end{figure*}

In practice, I drew $M=200$ realisations of the field on $\mathcal{S}^2$, that is, $200$ random maps. For each map~$m$, I computed the actual correlation function~$\xi(\theta; m)\neq \xi(\theta)$; then I drew $S=200$ random samples of $N=500$ directions, arranged the pairs of directions in $25$ equally populated bins, and computed the estimator~$\hat{\xi}_a(m, s)$ of the correlation function for each bin~$a$, map~$m$ and sample~$s$. Finally, I computed the covariance matrix as
\begin{multline}
\label{eq:sparsity_covariance_numerical}
\SCov(\hat{\xi}_a, \hat{\xi}_b)
\overset{\text{num}}{=}
\frac{1}{M} \sum_{m=1}^{M} \frac{1}{S}\sum_{s=1}^S
\pac{\hat{\xi}_a(m, s) - \xi(\bar{\theta}_a; m)}
\\ \times
\pac{\hat{\xi}_b(m, s) - \xi(\bar{\theta}_b; m)} ,
\end{multline}
where $\bar{\theta}_a$ is the mean separation of pairs in the $a$\textsuperscript{th} bin. The result is shown in the left panel of \cref{fig:test_sparsity_covariance}.

In the same figure, the central panel shows the theoretical prediction of \eqref{eq:SCov_result_Gaussian} for that model, which agrees very well with the numerical computation. Their difference, shown in the right panel, is due to the finiteness of the samples, $S, M<\infty$.

\subsection{Discussion}

It is interesting to examine the result of \cref{eq:SCov_result_Gaussian} in the case where the correlation function quickly decreases with angular separation. In that case, the sparsity contribution to the variance of $\hat{\xi}_a$ is indeed dominated by
\begin{equation}
\SCov(\hat{\xi}_a, \hat{\xi}_a)
\approx \frac{2}{N^2} \frac{\Omega}{\Omega_a} \, \xi^2(0)
= \frac{2}{P_a} \, \sigma_f^4\ ,
\end{equation}
where $\sigma_f$ denotes the standard deviation of $f$.

The interpretation of this result is quite straightforward, for $\hat{\xi}_a$ is the average of $f_i f_j$ over $P_a$ pairs. From one pair to the next, both $f_i$ and $f_j$ typically varies by $\sigma_f$, so that their product typically varies by $\sigma_f^2$; and since there are $P_a/2$ actual pairs in a bin, the standard deviation of their arithmetic mean is $\sigma_f^2/\sqrt{P_a/2}$.

The case discussed here, where the correlation function quickly decreases with the angular separation, is particularly relevant to cosmology. The presence of $\xi(0)$ in the sparsity contribution thus tends to boost it with respect to the cosmic contribution. As a rule of thumb, sparsity variance must be taken into account if the number of pairs~$P_a$ in a bin is not much larger than $\xi^2(0)/(\text{mean $\xi^2$ over $\footprint$})$.

\subsection{Comparison with shot noise}

The concept of sparsity covariance presented here may be reminiscent of shot noise. Shot noise emerges as one attempts to measure the macroscopic features of a physical quantity that is discrete at a more fundamental level. Pressure, in a fluid, is perturbed by discreteness of the fluid's constituents which is responsible for Brownian motion; electric current, in a conductor, is subject to fluctuations due to the discreteness of electrons; similarly, the intensity of a laser fluctuates due to the corpuscular nature of photons. In physical cosmology, shot noise typically appears as one aims to estimate the matter density contrast from galaxy number counts.

Although sparsity covariance encodes fluctuations that are related to the discreteness of a sample, I believe that it is conceptually distinct from shot noise. The key difference is that, contrary to, e.g., galaxy number counts, the general field $f$ considered in this section is not fundamentally discrete: it is a continuous field that is \emph{evaluated} in a discrete set of directions.

Having said that, the techniques employed in this article would also allow one to compute the contribution of shot noise in, e.g, galaxy-clustering or galaxy--galaxy lensing estimators. I refer the interested reader to \cite{Duboscq+2025} for a detailed illustration.

\section{Application to cosmic shear}
\label{sec:application_cosmic_shear}

Let us now examine how sparsity covariance manifests itself in the concrete case of cosmic shear, which is one of the key observables of modern cosmology~\citep{KiDS:2020suj, DES:2021vln, Euclid:2024yrr}.

\subsection{Reminder about cosmic shear}

This subsection is an executive summary of the relevant concepts and equations for cosmic shear; \correction{experts may skip to subsec.~\ref{subsection:full_covariance_cosmic_shear}}. For further details, I refer the reader to, e.g., \cite{Bartelmann:1999yn, Kilbinger:2014cea}.

\subsubsection{Weak gravitational lensing}

Gravity bends light rays. In the immediate vicinity of matter concentrations, this phenomenon can lead to significant distortions in the observed images of distant sources. Far from such concentrations, distortions are still present but typically unnoticeable by eye. This is the weak-lensing regime that applies to the vast majority of observed sources.

In particular, when observing a galaxy at cosmological distances, the deflection of light first implies that the galaxy is not observed where it actually is; second, and most importantly, the image of the galaxy is slightly distorted. This is because the image is composed of many rays that experience an inhomogeneous gravitational field as they propagate from the source to the observer, and hence are not deflected identically.

At leading order in the gradients of the gravitational field, the distortions of an image reduces to shear: a circular source would appear slightly elliptical. As a consequence, if we denote with $\eps_0\in\mathbb{C}$ the intrinsic ellipticity of a galaxy,\footnote{We use a complex number to encode both the amplitude and the orientation of the ellipticity. A popular choice is $\eps=(a-b)/(a+b) \exp(2\ii\psi)$, where $a, b$ are the semi-major and semi-minor axes of the ellipse, and $\psi$ the angle formed by the major axis with a fiducial direction~\citep{Bartelmann:1999yn}.} then its observed ellipticity reads
\begin{equation}
\eps = \eps_0 + \gamma \ ,
\end{equation}
where $\gamma\in\mathbb{C}$ represents the shear distortion due to weak gravitational lensing.

Technically speaking, shear is a spin-two field on the celestial sphere, which also depends on the redshift~$z_*$ of the source to which it is applied, $\gamma(\uvect{u}, z_*)$. It is convenient to express it in terms of a scalar potential~$\Psi(\uvect{u}, z_*)$ as
\begin{equation}
\label{eq:shear_derives_from_potential}
\gamma(\uvect{u}, z_*) = \frac{1}{2} \, \eth^2 \Psi(\uvect{u}, z_*) \ ,
\end{equation}
where $\eth$ is called the edth (or spin-raising) derivative, which may be seen as an angular derivative acting on the $\uvect{u}$ variable. See e.g. \cite{Castro:2005bg} for details.

In a spatially flat cosmological background with perturbations devoid of anisotropic stress, the lensing potential associated with those perturbations reads
\begin{equation}
\label{eq:lensing_potential}
\Psi(\uvect{u}, z_*)
=
2\int_0^{\chi_*} \frac{\dd\chi}{\chi} \; q(\chi; \chi_*) \,
\Phi(\eta_0-\chi, \chi, \uvect{u}) \ ,
\end{equation}
where $\chi$ denotes the comoving distance from the observer, so that $\chi_*=\chi(z_*)$ is the comoving distance to the source; $\Phi(\eta, \chi, \uvect{u})$ is the Bardeen potential, that is the gravitational potential produced by inhomogeneities in the distribution of matter in the Universe, evaluated at conformal time~$\eta$ and comoving distance~$\chi$ along a direction~$\uvect{u}$ in the sky; $\eta_0$ is conformal time today; and
\begin{equation}
q(\chi; \chi_*) \define \frac{\chi_* - \chi}{\chi_*}
\end{equation}
is a dimensionless integration kernel.

Together, \cref{eq:shear_derives_from_potential,eq:lensing_potential} show that $\gamma$ is sensitive to cosmological perturbations all along the line of sight from the source to the observer down its past light cone. Therefore, measurements of the weak-lensing shear tell us about the matter distribution in the Universe.

\subsubsection{Cosmic shear estimators}

For individual galaxies, it is impossible to distinguish the contribution of intrinsic ellipticity from the contribution of the weak-lensing shear. However, while galaxies are randomly oriented -- modulo intrinsic alignments, see e.g. \cite{Kirk:2015nma} -- shear exhibits correlations over large scales, which offers us the possibility to measure its effect in a statistical manner. In practice, this can be done using the two-point correlation function of $\eps$, so as to extract the two-point correlation function of $\gamma$. Since $\eps$ is a spin-two quantity, and since galaxies are distributed in redshift, the definition of such a correlation function is more subtle than in the scalar case of sec.~\ref{sec:sparsity_covariance}.

\begin{figure}[t]
\centering
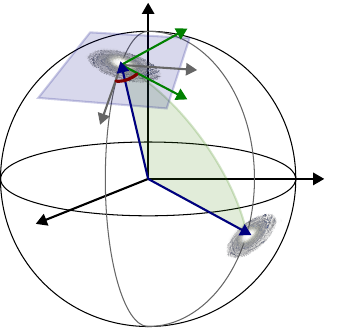
\caption{A pair of galaxies is observed in directions~$\uvect{u}_i, \uvect{u}_j$, forming an angle $\theta_{ij}$. This pair of directions is connected by a portion of great circle on $\mathcal{S}^2$, whose tangent vector oriented from $i$ to $j$ is $\uvect{t}_{ij}$. This vector, together with $\uvect{n}_{ij}=\uvect{u}_i\times \uvect{t}_{ij}$, forms a direct orthonormal basis rotated by an angle $\psi_{ij}$ with respect to the coordinate basis $(\uvect{e}_\theta, \uvect{e}_\ph)$. In the basis $(\uvect{t}_{ij}, \uvect{n}_{ij})$, the ellipticity~$\eps_i$ of the galaxy observed at $\uvect{u}_i$ can be decomposed into plus and cross components according to \cref{eq:ellipticity_plus_cross}.
}
\label{fig:cosmic_shear}
\end{figure}

The construction goes as follows (see \cref{fig:cosmic_shear}). For any pair $(ij)$ of galaxies, observed in directions~$(\uvect{u}_i, \uvect{u}_j)$ forming an angle $\theta_{ij}$, one first defines the components of galaxy ellipticities with respect to the axis connecting them. Consider the portion of great circle connecting $\uvect{u}_i$ to $\uvect{u}_j$, let $\uvect{t}_{ij}$ be the unit tangent vector oriented from $i$ to $j$, and $\uvect{n}_{ij}=\uvect{u}_i\times \uvect{t}_{ij}$. The set $(\uvect{t}_{ij}, \uvect{n}_{ij})$ forms a direct othonormal basis of the plane tangent to $\mathcal{S}^2$ at $\uvect{u}_i$, which is rotated by an angle $\psi_{ij}$ with respect to the basis~$(\uvect{e}_\theta, \uvect{e}_\ph)$ associated with polar coordinates. One then decomposes the ellipticity~$\eps_i$ of galaxy $i$ as
\begin{equation}
\label{eq:ellipticity_plus_cross}
\eps^+_{ij} + \ii  \eps^\times_{ij}
=
\eps_i \, \ex{-2\ii \psi_{ij}} \ .
\end{equation}
In the above, the ``plus'' and ``cross'' components,\footnote{In the weak-lensing literature, the ``plus'' component is more commonly referred to as ``tangential'' and denoted with $\eps\e{t}$. I personally find this terminology misleading for non-experts, since $\eps_+<0$ is radial rather than tangential. The ``plus/cross'' denomination is borrowed from the gravitational-wave literature~\citep{Maggiore:2007ulw}.}
$\eps^+_{ij},  \eps^\times_{ij}$, respectively represent the real and imaginary part of $\eps_i$ if the latter were expressed in the basis $(\uvect{t}_{ij}, \uvect{n}_{ij})$. The same decomposition can then be applied to galaxy $j$, that is, $\eps^+_{ji} + \ii  \eps^\times_{ji}= \eps_j \exp(-2\ii \psi_{ji})$. Note that $\psi_{ij}\neq \psi_{ji}$, although both coincide in the limit $\theta_{ij}\ll 1$ (flat-sky approximation).

Galaxies are distributed in three dimensions $(\uvect{u}_i, z_i)$. In a survey with a large number~$N$ of galaxies, relevant information can be extracted by dividing the sample into several redshift bins. Let $\redshiftbin_I$ be the $I$\textsuperscript{th} such bin -- in the following, Latin capital indices of the middle of the alphabet ($I, J, K, \ldots$) will refer to redshift bins -- and let $N_I=|\redshiftbin_I|$ be the number of galaxies in that bin.

We may then consider pairs of galaxies that belong to different redshift bins. Similarly to the set $\pairs_a$ defined in subsec.~\ref{subsec:estimator_correlation_function}, we define
\begin{equation}
\pairs_{aIJ} = \left\{ (ij) \; | \; z_i\in\redshiftbin_I, z_j\in\redshiftbin_J, \theta_{ij}\in[\theta_a, \theta_{a+1})\right\}
\end{equation}
as the set of pairs of galaxies $(ij)$ such that $i$ belongs to the $I$\textsuperscript{th} redshift bin, $j$ to the $J$\textsuperscript{th} redshift bin, and whose angular separation falls in the $a$\textsuperscript{th} angular bin. Note that when $I\neq J$, the set of pairs is not symmetric: if $(ij)\in\pairs_{aIJ}$, then $(ji)\in\pairs_{aJI}\neq\pairs_{aIJ}$.

Equipped with such definitions, we can finally build estimators for the correlation functions of galaxy ellipticity,
\begin{equation}
\label{eq:estimator_xi_pm}
\hat{\xi}_{a IJ}^\pm
\define
\frac{1}{P_{aIJ}}
\sum_{(ij) \in \pairs_{a IJ}}
\pa{\eps^+_{ij} \eps^+_{ji} \pm \eps^\times_{ij} \eps^\times_{ji} } ,
\end{equation}
with $P_{aIJ}=|\pairs_{aIJ}|$. The fact that two different correlation functions, $\xi^+$ and $\xi^-$, can be defined, is related to the spin-two nature of $\eps$. Geometrically speaking, $\xi^+$ captures small-scale correlations, when pairs of galaxies are subject to the same shear field, $\gamma_i \approx \gamma_j$ and hence get aligned along the same direction. On the other hand, $\xi^-$ captures the tangential alignment of galaxies all around matter overdensities, which translates as $\gamma_{ij}\approx \gamma_{ij}^*$; it is therefore more sensitive to correlations on larger scales.

\subsubsection{Expectation value of the estimators}

Just like in sec.~\ref{sec:sparsity_covariance}, one computes the expectation value of the estimator by first applying cosmic averaging, and then sample averaging. When taking the cosmic average of the summand in \cref{eq:estimator_xi_pm}, $\eps^+_{ij} \eps^+_{ji} \pm \eps^\times_{ij} \eps^\times_{ji}$, the contribution of intrinsic ellipticities vanishes if one assumes their orientations to be uncorrelated for $i\neq j$. Thus,
\begin{align}
\cev{ \eps^+_{ij} \eps^+_{ji} \pm \eps^\times_{ij} \eps^\times_{ji} }
&= \cev{ \gamma^+_{ij} \gamma^+_{ji} \pm \gamma^\times_{ij} \gamma^\times_{ji} }
\\
&=
\xi^\pm_{ij}
\define
\xi^\pm(\theta_{ij}; z_i, z_j) \ ,
\end{align}
where $\xi^\pm(\theta, z_1, z_2)$ are the correlation functions of shear only, for two lines of sights ending at redshifts $z_1, z_2$ and separated by an angle $\theta$.

The theoretical expression of $\xi^\pm(\theta, z_1, z_2)$ is more conveniently established in harmonic space -- see e.g. \cite{Kilbinger:2017lvu} for details. As one decomposes the shear field into spin-weight spherical harmonics, the correlation functions read
\begin{equation}
\xi^\pm(\theta; z_1, z_2)
=
\frac{1}{4\pi} \sum_{\ell=2}^\infty (2\ell+1) d^\ell_{\pm 2 2}(\theta) \,
C_\ell(z_1, z_2)
\end{equation}
in terms of the shear angular power spectrum~$C_\ell(z_1, z_2)$, where $d^\ell_{s_1 s_2}(\theta)$ are elements of the Wigner $d$-matrices. In the flat-sky approximation, i.e. for $\theta\ll 1$, these matrices can be expressed in terms of Bessel functions as $d^\ell_{2 2}(\theta)\approx J_0(\ell\theta)$ and $d^\ell_{-2 2}(\theta)\approx J_4(\ell\theta)$, and $\ell$ may be treated as a continuous variable to get
\begin{equation}
\xi^\pm(\theta; z_1, z_2)
\approx
\frac{1}{2\pi}
\int_0^\infty \dd\ell \; \ell J_{0/4}(\ell\theta) \, C_\ell(z_1, z_2) \ .
\end{equation}

The advantage of the angular power spectrum~$C_\ell$ is that it is more easily connected to that of the lensing potential, $C^\Psi_\ell$. \Cref{eq:shear_derives_from_potential} indeed implies
\begin{equation}
C_\ell(z_1, z_2)
=
\frac{1}{4} \frac{(\ell+2)!}{(\ell-2)!} \, C^\Psi_\ell(z_1, z_2)
\end{equation}
which, together with \cref{eq:lensing_potential}, Poisson's equation and Limber's approximation, eventually yields
\begin{multline}
C_\ell(z_1, z_2)
\approx
\pa{\frac{3}{2} \, \frac{H_0^2}{c^2}\Omega\e{m}}^2
\int_0^\infty
\dd\chi \; [1+z(\chi)]^2 \\
\times q(\chi; \chi_1)  q(\chi; \chi_2)
P\e{m}\pac{z(\chi), \frac{\ell+1/2}{\chi}} ,
\end{multline}
where $P\e{m}(z, k)$ is the matter power spectrum at redshift~$z$ and comoving scale~$k$; $H_0$ the Hubble constant; $c$ the speed of light in vacuum, and $\Omega\e{m}$ the matter density today normalised by the cosmic critical density.

The second step consists in averaging over samples. The procedure is essentially the same as in subsubsec.~\ref{subsubsec:expectation_value_estimator}, except that galaxies are distributed in three dimensions. In practice, this means that one must average over all possible redshifts $z_i, z_j$ on top of angular positions for pairs $(ij)\in\pairs_{aIJ}$. Let $p_I(z)$ denote the probability density function of redshift in the $I$\textsuperscript{th} bin. Assuming the latter to be independent of the angular positions of galaxies, it is straightforward to get
\begin{multline}
\sev{\cev{\hat{\xi}_{aIJ}^\pm}}
= 
\xi^\pm_{\sev{12}_{aIJ}}
\define
\int_0^\infty \dd z_1 \; p_I(z_1)
\int_0^\infty \dd z_2 \; p_J(z_2) \\
\times
\int_{\cos\theta_{a+1}}^{\cos\theta_a}
\frac{\dd\cos\theta}{\cos\theta_a - \cos\theta_{a+1}} \;
\xi^\pm(\theta; z_1, z_2) \ .
\end{multline}
In the above, I generalised the notation $X_{\sev{12}_a}$ of sec.~\ref{sec:sparsity_covariance} into $X_{\sev{12}_{aIJ}}$ by adding redshift integration.

\paragraph{Remark} In the cosmic-shear literature, it customary to integrate over redshift much earlier in the calculation. In \cite{Kilbinger:2017lvu} for instance, this is done in the very first equation, by defining an effective lensing potential, and hence an effective shear, that is already marginalised over the galaxy sample's redshifts, and hence only depends on the direction of observation~$\uvect{u}$. This leads to the binned angular power spectra
\begin{align}
C^{IJ}_\ell
&\define
\int_0^\infty \dd z_1 \; p_I(z_1)
\int_0^\infty \dd z_2 \; p_J(z_2) \;
C_\ell^\gamma
\\
&=
\pa{\frac{3}{2} \, H_0^2\Omega\e{m}}^2
\int_0^\infty
\dd\chi \; [1+z(\chi)]^2
\nonumber \\&\quad
\times q_I(\chi)  q_J(\chi)
P\e{m}\pac{z(\chi), \frac{\ell+1/2}{\chi}} ,
\end{align}
with
\begin{equation}
q_I(\chi) = \int_0^\infty \dd z \; p_I(z) \, q[\chi; \chi_*(z)]
\end{equation}
Albeit technically correct, this approach has, in my opinion, the pedagogical disadvantage of putting the variables $z$ and $\uvect{u}$ on different footings. This may be misleading, in particular, when dealing with sample averaging in the calculation of the covariance matrix.

\subsection{Full covariance matrix of cosmic shear}
\label{subsection:full_covariance_cosmic_shear}

The covariance matrix is defined as
\begin{multline}
\label{eq:covariance_cosmic_shear_def}
\Cov\pa{\hat{\xi}_{aIJ}^\sigma, \hat{\xi}_{bKL}^\eta}
\\
=
\sev{\cev{\hat{\xi}_{aIJ}^\sigma \hat{\xi}_{bKL}^\eta}}
-
\sev{\cev{\hat{\xi}_{aIJ}^\sigma}} \sev{\cev{\hat{\xi}_{bKL}^\eta}} ,
\end{multline}
with $\sigma, \eta=\pm$. It was first computed by \cite{2002A&A...396....1S}; 
a more modern and comprehensive treatment is presented by \cite{Reischke:2024fvk}, whose results are implemented in the \OneCov code,\footnote{\url{https://onecovariance.readthedocs.io}} but still apparently miss the sparsity term.

I propose an updated derivation here. Following the same rationale as in sec.~\ref{sec:sparsity_covariance}, I shall not attempt to calculate sparsity covariance on its own, but rather extract it from the full final expression. However, since the equations will quickly become heavy, and since most of the relevant expressions were already presented in \cite{2002A&A...396....1S}, I shall mostly describe the general method and give the final results in the main text; further details on the intermediate steps can be found in appendix~\ref{appendix:details_derivation_covariance}.

The expectation values of $\hat{\xi}_{aIJ}^\sigma, \hat{\xi}_{bKL}^\eta$ that appear in \cref{eq:covariance_cosmic_shear_def} have already be taken care of, so the core of the calculation lies in the quadratic term.

\subsubsection{Cosmic averaging}

Similarly to the simple scalar case of sec.~\ref{sec:sparsity_covariance}, the cosmic expectation value of the quadratic term reads
\begin{equation}
\label{eq:cosmic_average_quadratic_term_cosmic_shear}
\cev{\hat{\xi}_{aIJ}^\sigma \hat{\xi}_{bKL}^\eta}
=
\frac{1}{P_{aIJ} P_{bKL}} \sum_{(ij) \in \pairs_{a IJ}} \sum_{(kl) \in \pairs_{b KL}}
\Xi^{\sigma\eta}_{ijkl} \ ,
\end{equation}
with the four-point correlation function
\begin{equation}
\Xi^{\sigma\eta}_{ijkl}
\define
\cev{
\pa{ \eps^+_{ij}\eps^+_{ji} + \sigma \eps^\times_{ij}\eps^\times_{ji} }
\pa{ \eps^+_{kl}\eps^+_{lk} + \eta \eps^\times_{kl}\eps^\times_{lk} }
} .
\end{equation}

The explicit expression of $\Xi^{\sigma\eta}_{ijkl}$ [see \cref{eq:Xi_++,eq:Xi_--,eq:Xi_+-}] is obtained by expanding its definition using $\eps=\eps_0 +\gamma$. Since $\eps_0, \gamma$ are treated as independent, Gaussian random fields, Isserlis's theorem implies that $\Xi$ is a sum of terms of the form $\zeta_1\times \zeta_2$, where $\zeta_1, \zeta_2\sim \cev{\gamma\gamma}$ or $\cev{\eps_0\eps_0}$. On the one hand, $\cev{\gamma\gamma}$ can be expressed in terms of $\xi^\pm$; on the other hand, $\eps_0$ is assumed to be delta-correlated (no intrinsic alignments),
\begin{equation}
\cev{\eps_{0i} \eps_{0j}} = 0 \ ,
\quad
\cev{\eps_{0i} \eps_{0j}^*} = \sigma_i^2 \delta_{ij} \ .
\end{equation}
The mean square of intrinsic ellipticity~$\sigma^2$ considered here matches the convention of \cite{2002A&A...396....1S}. It is, however, two times larger than the $\sigma^2_1$ used in \cite{Reischke:2024fvk} and \OneCov, which is the mean square of the real or imaginary part of $\eps_0$.

\subsubsection{Sample averaging}

The second step consists in taking the sample average of \cref{eq:cosmic_average_quadratic_term_cosmic_shear}. This is where the present derivation begins to depart from \cite{2002A&A...396....1S}. In the latter, sample averaging is taken as a marginalisation over the positions of galaxies $i, j, k, l$, which are considered independent. But as we have seen in subsubsec.~\ref{subsubsec:mathematical_origin_sparsity_covariance}, this does not work when two indices happen to be identical; those cases precisely lead to sparsity covariance.

Just like in subsubsec.~\ref{subsubsec:mathematical_origin_sparsity_covariance}, one may isolate, in the double sum of \cref{eq:cosmic_average_quadratic_term_cosmic_shear}, the terms that have one index, or two indices, in common. The only difference is that, here, the four possibilities for one common index $(i=k, i=l, j=k, j=l)$ are not equivalent because, e.g., $i=k$ requires the associated galaxies to belong to the same bin ($I=K$); likewise for the two possibilities for two common indices [$(ij)=(kl)$ and $(ij)=(lk)$]. Apart from that difference, the logic is the same, and the full covariance reads
\begin{equation}
\Cov
=
\CCov + \NCov + \SCov \ ,
\end{equation}
where, analogously to \cref{eq:cosmic_covariance_semi_explicit},
\begin{equation}
\label{eq:cosmic_covariance_cosmic_shear}
\CCov\pa{\hat{\xi}_{aIJ}^\sigma, \hat{\xi}_{bKL}^\eta}
=
\Xi^{\sigma\eta}_{\sev{12}_{aIJ} \sev{34}_{bKL}} - \xi^\sigma_{\sev{12}_{aIJ}} \xi^\eta_{\sev{34}_{bKL}}
\end{equation}
is independent of the number of galaxies in the sample; it is therefore identified with cosmic covariance. Its explicit expression, already established in the literature, is reproduced in \cref{eq:cosmic_covariance_shear_++,eq:cosmic_covariance_shear_--,eq:cosmic_covariance_shear_+-}.

As for the other two components, I find
\begin{widetext}
\begin{multline}
\label{eq:noise_sparsity_covariance_cosmic_shear}
(\NCov+\SCov)\pa{\hat{\xi}_{aIJ}^\sigma, \hat{\xi}_{bKL}^\eta}
=
\frac{\delta_{IK}}{N_I}
\pa{
\Xi^{\sigma\eta}_{\sev{12}_{aIJ} \sev{14}_{bIL}} - \Xi^{\sigma\eta}_{\sev{12}_{aIJ} \sev{34}_{bKL}}
}
+ \frac{\delta_{IL}}{N_I}
\pa{
\Xi^{\sigma\eta}_{\sev{12}_{aIJ} \sev{31}_{bKI}} - \Xi^{\sigma\eta}_{\sev{12}_{aIJ} \sev{34}_{bKL}}
}
\\
+ \frac{\delta_{JK}}{N_J}
\pa{
\Xi^{\sigma\eta}_{\sev{12}_{aIJ} \sev{24}_{bJL}} - \Xi^{\sigma\eta}_{\sev{12}_{aIJ} \sev{34}_{bKL}}
}
+ \frac{\delta_{JL}}{N_J}
\pa{
\Xi^{\sigma\eta}_{\sev{12}_{aIJ} \sev{32}_{bKJ}} - \Xi^{\sigma\eta}_{\sev{12}_{aIJ} \sev{34}_{bKL}}
}
\\
+
\frac{\delta_{ab} \delta_{IK}\delta_{JL}}{P_{aIJ}}
\pa{ \Xi^{\sigma\eta}_{\sev{12}_{aIJ} \sev{12}_{aIJ}}  - \Xi^{\sigma\eta}_{\sev{12}_{aIJ} \sev{34}_{bKL}} }
+
\frac{\delta_{ab} \delta_{IL}\delta_{JK}}{P_{aIJ}}
\pa{ \Xi^{\sigma\eta}_{\sev{12}_{aIJ} \sev{21}_{aJI}}  - \Xi^{\sigma\eta}_{\sev{12}_{aIJ} \sev{34}_{bKL}} }
.
\end{multline}
\end{widetext}
The first two lines of \cref{eq:noise_sparsity_covariance_cosmic_shear} stem from the terms, in the sum of \cref{eq:cosmic_average_quadratic_term_cosmic_shear}, that have one index in common, whilst the third line comes from terms with two indices in common. In the case where there is only one redshift bin (hence $I=J=K=L$), one recovers \cref{eq:sparsity_covariance_semi_explicit}.

\Cref{eq:noise_sparsity_covariance_cosmic_shear} encodes both the contributions of intrinsic ellipticity (the so-called shape noise), and sparsity covariances. I choose not to separate them here, for reasons that will soon become clear. In the pseudo-full-sky and flat-sky ($\psi_{ij}=\psi_{ji}$) approximations, their explicit expressions are
\begin{widetext}
\begin{align}
\label{eq:sparsity+noise_++}
&\pa{\NCov + \new{\SCov}}\pa{\hat{\xi}_{aIJ}^+, \hat{\xi}^+_{bKL}}
\nonumber \\
&=
\frac{1}{2}
\frac{\delta_{IK}}{N_I}
\pac{
\sigma_I^2 + \new{\xi_I^+(0)}
}
\int_{\annulus_a} \frac{\dd^2\uvect{u}_2}{\Omega_a}
\int_{\annulus_b} \frac{\dd^2\uvect{u}_4}{\Omega_b} \;
\xi^+_{JL}(\theta_{24})
+ \text{$3$ permutations $(I\leftrightarrow J$ or $K\leftrightarrow L)$}
\nonumber\\
&\quad +
\frac{1}{2}
\frac{\delta_{ab} \delta_{IK} \delta_{JL}}{P_{aIJ}}
\pac{\sigma_I^2 + \new{\xi_I^+(0)}}
\pac{\sigma_J^2 + \new{\xi_J^+(0)}}
+ \text{$1$ permutation $(I\leftrightarrow J$ and $K\leftrightarrow L)$} \ ,
\\
\label{eq:sparsity+noise_--}
&\pa{\NCov + \new{\SCov}}\pa{\hat{\xi}_{aIJ}^-, \hat{\xi}^-_{bKL}}
\nonumber \\
&=
\frac{1}{2}
\frac{\delta_{IK}}{N_I}
\pac{
\sigma_I^2 + \new{\xi_I^+(0)}
}
\int_{\annulus_a} \frac{\dd^2\uvect{u}_2}{\Omega_a}
\int_{\annulus_b} \frac{\dd^2\uvect{u}_4}{\Omega_b} \;
\xi^+_{JL}(\theta_{24})\, \cos4(\ph_2-\ph_4)
+ \text{$3$ permutations $(I\leftrightarrow J$ or $K\leftrightarrow L)$}
\nonumber\\
&\quad +
\frac{1}{2}
\frac{\delta_{ab} \delta_{IK} \delta_{JL}}{P_{aIJ}}
\pac{\sigma_I^2 + \new{\xi_I^+(0)}}
\pac{\sigma_J^2 + \new{\xi_J^+(0)}}
+ \text{$1$ permutation $(I\leftrightarrow J$ and $K\leftrightarrow L)$} \ ,
\end{align}
\begin{align}
\label{eq:sparsity+noise_+-}
&\pa{\NCov + \new{\SCov}}\pa{\hat{\xi}_{aIJ}^+, \hat{\xi}^-_{bKL}}
\nonumber \\
&=
\frac{1}{2}
\frac{\delta_{IK}}{N_I}
\pac{
\sigma_I^2 + \new{\xi_I^+(0)}
}
\int_{\annulus_a} \frac{\dd^2\uvect{u}_2}{\Omega_a}
\int_{\annulus_b} \frac{\dd^2\uvect{u}_4}{\Omega_b} \;
\xi_{JL}^-(\theta_{24})\, \cos4 (\psi_{24}-\ph_2)
+ \text{$3$ permutations $(I\leftrightarrow J$ or $K\leftrightarrow L)$} \ ,
\end{align}
\end{widetext}
up to terms $\sim\CCov/N$ that are negligible in practice -- see \cref{eq:sparsity_noise_covariance_shear_++_complete,eq:sparsity_noise_covariance_shear_--_complete,eq:sparsity_noise_covariance_shear_+-_complete} for the full expressions. I have indicated in green the \new{new terms with respect to the cosmic-shear literature}. The angles denoted with $\ph$ are the usual azimuthal polar coordinate of a direction $\uvect{u}(\theta, \ph)\in\mathcal{S}^2$. I have also used the shorthand notation
\begin{align}
\xi^\pm_{IJ}(\theta)
&\define
\int_0^\infty\dd z_1 \; p_I(z_1) \int_0^\infty \dd z_2 \; p_J(z_2) \, \xi^\pm(\theta; z_1, z_2) \ ,
\\
\xi^\pm_{I}(\theta)
&\define
\int_0^\infty\dd z \; p_I(z) \, \xi^\pm(\theta; z, z) \ ,
\end{align}
while $\sigma^2_I$ denotes the mean square of intrinsic ellipticities in the $I$\textsuperscript{th} redshift bin,
\begin{equation}
\sigma_I^2
\define
\frac{1}{N_I} \sum_{i\in\redshiftbin_I} \sigma_i^2 
\approx
\frac{1}{N_I} \sum_{i\in\redshiftbin_I} |\eps_{0, i}|^2 \ .
\end{equation}

\subsection{Sparsity covariance acts as an extra shape noise}

\Cref{eq:sparsity+noise_++,eq:sparsity+noise_--,eq:sparsity+noise_+-} are written in a way that emphasises the fact that the main contribution of sparsity to the covariance matrix of cosmic shear is equivalent to enhancing shape noise as
\begin{equation}
\sigma_I^2 \to \sigma_I^2 + \xi^+_I(0) \ . 
\end{equation}
This quantity has a very clear physical interpretation: since $\xi^+_I(0)$ is the mean square of shear in the $I$\textsuperscript{th} bin, $\sigma_I^2 + \xi^+_I(0)$ is simply the mean square of the \emph{apparent} ellipticity of galaxies in that bin.

Therefore, \emph{sparsity covariance is already accounted for in cosmic-shear analyses}, because in practice the shape noise amplitude is estimated from the apparent ellipticity of galaxies, \correction{their intrinsic ellipticity being unobservable.}

\begin{figure}[t]
\centering
\includegraphics[width=\columnwidth]{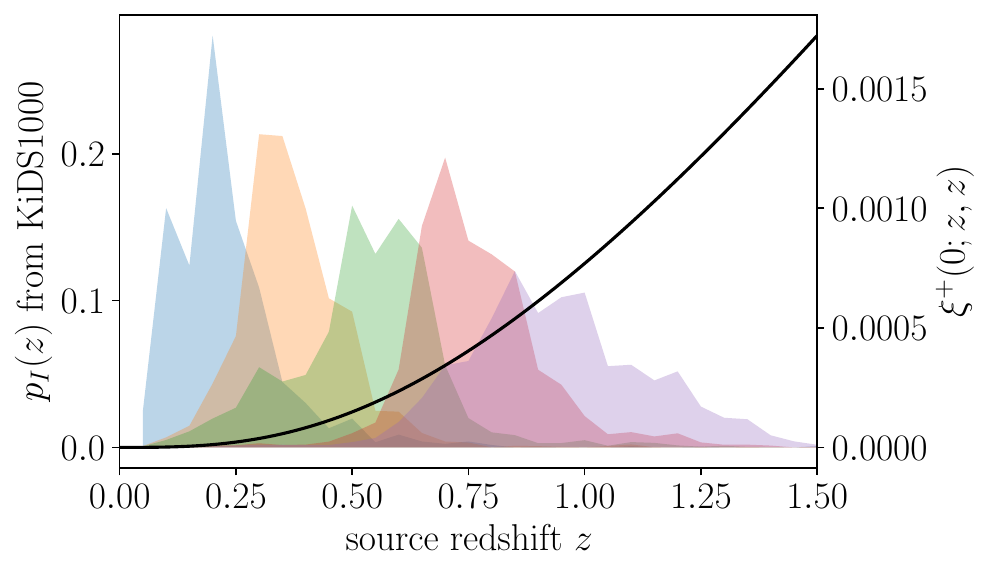}
\caption{\correction{Mean square of shear~$\xi^+(0; z, z)$  for a source at redshift~$z$ (black line), together with the redshift distributions of the five tomographic bins of KiDS1000 (semi-transparent histograms).} \notebook}
\label{fig:variance_shear_ellipticity}
\end{figure}

\correction{Even if we could estimate the mean square of the sole intrinsic ellipticity~$\sigma_I^2$, the contribution of shear~$\xi^+_I(0)$ would represent a sub-percent correction in the range of redshift probed by weak-lensing surveys. For illustration, \cref{fig:variance_shear_ellipticity} indicates the redshift distributions~$p_I(z)$ of the five tomographic bins of the KiloDegree Survey [KiDS1000, \cite{KiDS:2020suj}], which extends up to $z\approx 1.5$, together with $\xi^+(0; z, z)$ estimated for a \textit{Planck}-2018 cosmology~\citep{Planck:2018vyg} using \textsc{camb}.\footnote{\url{https://camb.readthedocs.io}} The latter} is found to be well fit for $z<2$ by\footnote{The exact value of $\xi^+(0; z, z)$ is actually difficult to estimate with precision, because it is sensitive to the power spectrum all the way down to galactic scales~\citep{Fleury:2018cro} -- see also \cite{DES:2025omh} for a recent discussion. The estimate proposed here relies on a power-law extrapolation of the non-linear power spectrum generated by \textsc{camb} and \textsc{halofit}.} 
\begin{equation}
\correction{\xi^+(0; z, z)}
= \pac{(1 + a z^b)^c-1}^2 \ ,
\end{equation}
with $(a, b, c)=(1.178, 1.519, 0.035)$. \correction{This is at least two order of magnitudes smaller than the typical squared ellipticity $\sigma^2 \sim 2 \times 0.3^2 \sim 0.2$ of a galaxy. A more precise comparison of the mean square of shear and intrinsic ellipticity in each of the KiDS1000 tomographic bins is provided in \cref{tab:comparison_sigma2_xi0_KiDS1000}, which shows that $\xi^+_I(0)$ indeed represents a sub-percent fraction of $\sigma_I^2$.}

\begin{table}[h]
\caption{\correction{Comparison of the intrinsic ellipticity dispersion~$\sigma_I^2$ and the mean square of shear~$\xi^+_I(0)$ in each tomographic bin~$I$ of KiDS1000.}}
\label{tab:comparison_sigma2_xi0_KiDS1000}
\centering
\begin{tabular}{cccc}
\hline
$I$ & $\sigma_I^2$ &  $\xi_I^+(0)$ & $\xi_I^+(0)/\sigma_I^2$\\
\hline
1 & 0.146  	& $4.00\times 10^{-5}$	& 0.027\% \\
2 & 0.133  	& $9.06\times 10^{-5}$ & 0.068\%\\
3 & 0.149  	& $2.13\times 10^{-4}$ & 0.14\%\\
4 & 0.129  	& $4.44\times 10^{-4}$ & 0.34\%\\
5 & 0.145		& $7.37\times 10^{-4}$ & 0.51\%\\
\hline
\end{tabular}
\end{table}

\begin{figure*}[t]
\centering
\includegraphics[width=\columnwidth]{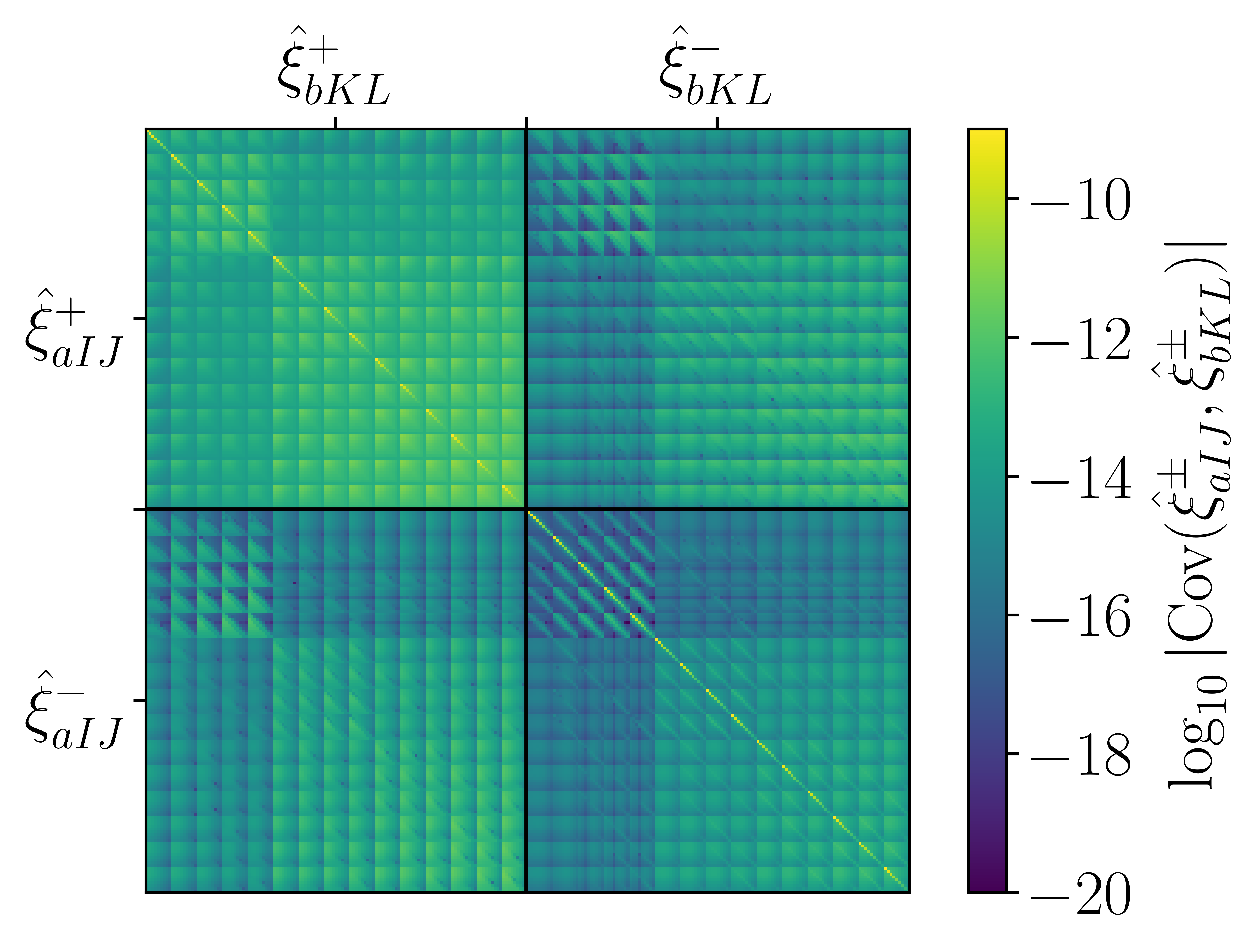}
\hfill
\includegraphics[width=\columnwidth]{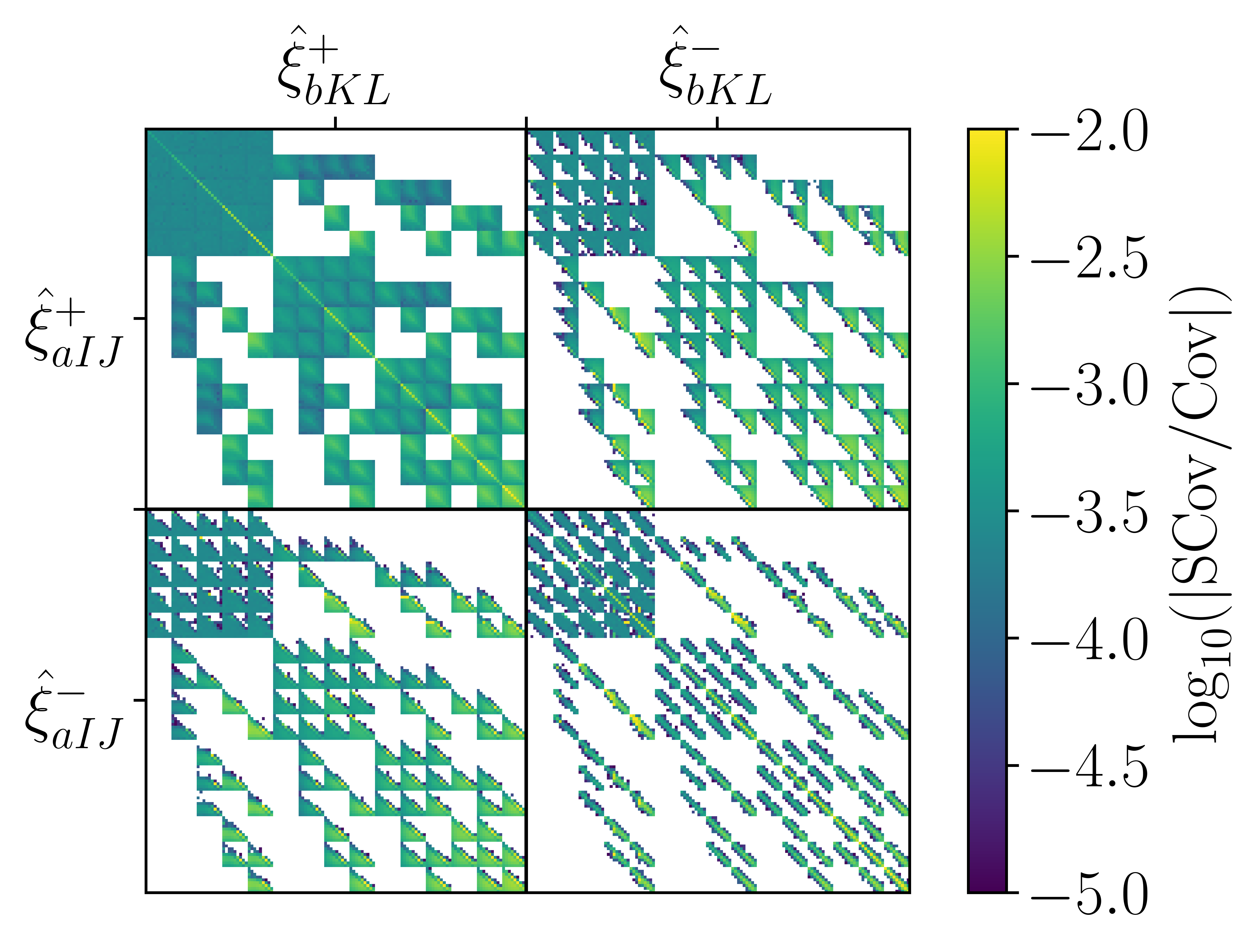}
\caption{\correction{\textit{Left panel}: covariance matrix of KiDS1000, computed with \OneCov. \textit{Right panel}: relative contribution of sparsity in that covariance matrix. Empty pixels indicate a contribution of sparsity that is either exactly zero (due to the Kr\"onecker deltas in the expression of $\SCov$) or below machine precision. For each panel, the matrices are organised as follows. There are 5 tomographic bins represented by indices $I, J, K, L$; and 9 angular bins indices represented by indices $a, b$. Within each of the main four blocks ($++, +-, -+, --$), the various indices run in the order $a, J, I$, with $J\leq I$ for rows, and $b, L, K$, with $L\leq K$ for columns. For example, successive rows correspond to $aIJ=111, \ldots, 911, 112, \ldots 912, \ldots [\text{$J$ runs from 1 to 5}] \ldots 915, 122, \ldots, 922, 123, \ldots 923, \text{etc.}$ \notebook}}
\label{fig:Cov_KiDS1000}
\end{figure*}

\correction{For the sake of completeness, \cref{fig:Cov_KiDS1000} depicts the full covariance matrix of KiDS1000 in real space (left panel), with 5 tomographic bins and 9 bins of angular separation, together with the relative contribution of sparsity (right panel). The covariance matrices were produced using the \OneCov\xspace code, assuming Gaussian fields for simplicity. In order to evaluate the contribution of sparsity, I ran \OneCov\xspace twice, with different values for the \texttt{ellipticity\_dispersion} parameter. In the first run, this parameter was set by the mean square of galaxy ellipticities as observed by KiDS,
\begin{equation}
\texttt{ellipticity\_dispersion}[I] = \sqrt{\frac{\sigma_I^2 + \xi_I^2(0)}{2}} \ ,
\end{equation}
with the values of \cref{tab:comparison_sigma2_xi0_KiDS1000}; the resulting covariance matrix, $\Cov_1$, thus accounts for sparsity. In the second run, I used $\texttt{ellipticity\_dispersion}[I]=\sigma_I/\sqrt{2}$, thereby producing a covariance matrix~$\Cov_2$ where sparsity is neglected. I defined sparsity covariance as
\begin{equation}
\SCov \define \Cov_1 - \Cov_2 \ .
\end{equation}
The two covariance matrices, as well as the configuration files used to produce them with \OneCov\xspace can be found on the GitHub repository linked to this article.

The right panel of \cref{fig:Cov_KiDS1000} shows that the relative contribution of sparsity in the KiDS1000 covariance matrix is at most on the order of $1\%$, as expected from \cref{tab:comparison_sigma2_xi0_KiDS1000}. The empty (white) pixels indicate that $\SCov$ is either exactly zero, due to the Kr\"onecker symbols appearing in \cref{eq:sparsity+noise_++,eq:sparsity+noise_+-,eq:sparsity+noise_--}, or below machine precision.
}



\section{Conclusion}
\label{sec:conclusion}

When estimating the statistical properties of a random field from a finite, discrete sample of measurements, the result depends on the sample at hand. Sparsity covariance quantifies the uncertainty specifically due to the discreteness of the sample. It thus differs from sample (or cosmic) covariance, which is due to its finite extension.

In this article, motivated by questions relevant to cosmology, I have mathematically defined sparsity covariance and shown how to evaluate it, in the context of measurements of the two-point correlation function of a random field observed in random directions of the sky. I first treated the case of a generic scalar field $f:\mathcal{S}^2\to\mathbb{R}$, \correction{observed without noise}, in order to convey the general idea. The result was validated with a numerical experiment. \correction{As a rule of thumb, sparsity covariance is $\sim \xi^2(0)/P$, where $P$ is the number of pairs of directions used to estimate $\xi(\theta)$. For comparison, cosmic covariance is typically given by the celestial average of $\xi^2(\theta)$.}

I then turned to the more concrete case of cosmic shear, where I showed that sparsity covariance essentially acts as an extra shape noise. Specifically, accounting for the dominant contribution of sparsity covariance is equivalent to replacing, in the expression of shape noise, the mean square of the intrinsic ellipticity of galaxies, $\cev[1]{|\eps_0|^2}$, by the mean square of their apparent ellipticity, $\cev[1]{|\eps|^2}=\cev[1]{|\eps_0|^2}+\cev[1]{|\gamma|^2}$, which accounts for weak lensing. As it turns out, since shape noise is, in practice, estimated from apparent ellipticities, sparsity covariance is already taken into account in standard cosmic shear analyses. If it were not, the associated error on the covariance matrix would remain negligible anyway, since $\cev[1]{|\gamma|^2}/\cev[1]{|\eps_0|^2}<1\%$.

Things would be different if the signal (here $\gamma$) were comparable or stronger than the noise (here $\eps_0$). More generally, the analysis presented here reveals that sparsity covariance must be taken into account when dealing with measurements whose individual signal-to-noise ratio is comparable, or larger, than unity. It will be the case, e.g., for cosmological applications of the line-of-sight shear of strong lenses~\citep{Hogg:2022ycw, Fleury:2024sgq}.

\section*{Acknowledgements}

I thank Théo Duboscq, Daniel Johnson, Julien Larena, Cyril Pitrou and Robert Reischke for useful discussions and for their feedback on the draft. Many thanks to Robert Reischke and Angus Wright for their assistance with \OneCov. Finally, I am very grateful to Natalie Hogg for helping me improve the written English of this article. This work is supported by the French \emph{Agence Nationale de la Recherche} through the ELROND project (ANR-23-CE31-0002).

\begin{widetext}
\appendix

\section{Details on the derivation of the covariance matrix of cosmic shear}
\label{appendix:details_derivation_covariance}

This appendix provides further details on the derivation of the covariance matrix of cosmic shear.

\subsection{Expression of $\Xi^{\sigma\eta}_{ijkl}$}

The first technical step consists in expressing the four-point correlation
\begin{equation}
\label{eq:definition_Xi_ijkl}
\Xi^{\sigma\eta}_{ijkl}
\define
\cev{
\pa{ \eps^+_{ij}\eps^+_{ji} + \sigma \eps^\times_{ij}\eps^\times_{ji} }
\pa{ \eps^+_{kl}\eps^+_{lk} + \eta \eps^\times_{kl}\eps^\times_{lk} }
} .
\end{equation}
in terms of the building blocks $\xi^\pm_{ij} \define \cev{\gamma^+_{ij}\gamma^+_{ji}} \pm  \cev{\gamma^\times_{ij}\gamma^\times_{ji}}$, and the mean square of $\eps_0$, $\cev{\eps_{0i}\eps_{0j}^*} = \sigma_i^2$. Note that intrinsic ellipticity can actually be treated just like shear with the substitution $\xi^+_{ij}\to\sigma_i^2\delta_{ij}$ and  $\xi^-_{ij}\to 0$.

Using the definition of the plus and cross components of ellipticity,
\begin{align}
\eps_i &= (\eps_{ij}^+ + \ii\eps_{ij}^\times)\ex{2\ii\psi_{ij}} = (\eps_{ik}^+ + \ii\eps_{ik}^\times)\ex{2\ii\psi_{ik}} \ ,\\
\eps_k &= (\eps_{kl}^+ + \ii\eps_{kl}^\times)\ex{2\ii\psi_{kl}} = (\eps_{ki}^+ + \ii\eps_{ki}^\times)\ex{2\ii\psi_{ki}} \ ,
\end{align}
it is straightforward to get
\begin{align}
\cev{\eps_{+ij}\eps_{+kl}}
&= \frac{1}{2}
    \pac{
        \pa{\xi^+_{ik} + \sigma_i^2\delta_{ik}} \cos2(\psi_{ik} - \psi_{ij} - \psi_{ki} + \psi_{kl})
        + \xi^-_{ik} \cos2(\psi_{ik} - \psi_{ij} + \psi_{ki} - \psi_{kl})
    },
\\
\cev{\eps_{\times ij}\eps_{\times kl}}
&= \frac{1}{2}
    \pac{
        \pa{\xi^+_{ik} + \sigma_i^2\delta_{ik}} \cos2(\psi_{ik} - \psi_{ij} - \psi_{ki} + \psi_{kl})
        - \xi^-_{ik} \cos2(\psi_{ik} - \psi_{ij} + \psi_{ki} - \psi_{kl})
    },
\\
\cev{\eps_{+ ij}\eps_{\times kl}}
&= \frac{1}{2}
    \pac{
        - \pa{\xi^+_{ik} + \sigma_i^2\delta_{ik}} \sin2(\psi_{ik} - \psi_{ij} - \psi_{ki} + \psi_{kl})
        + \xi^-_{ik} \sin2(\psi_{ik} - \psi_{ij} + \psi_{ki} - \psi_{kl})
    }.
\end{align}
Then, in the flat-sky limit ($\psi_{ij}=\psi_{ji}$), after a tedious expansion of \cref{eq:definition_Xi_ijkl} in terms of the above and the application Isserlis's theorem, I find
\begin{align}
\label{eq:Xi_++}
\Xi^{++}_{ijkl}
&=
\xi^+_{ij} \xi^+_{kl} +
\frac{1}{2} \pac{
\pa{\xi^+_{ik} + \sigma_i^2 \delta_{ik}} \pa[2]{\xi^+_{jl} + \sigma_j^2 \delta_{jl}}
+ \xi^-_{ik} \xi^-_{jl} \cos4(\psi_{ik}-\psi_{jl})
+ (k\leftrightarrow l)}
,
\\
\label{eq:Xi_--}
\Xi^{--}_{ijkl}
&= \xi^-_{ij} \xi^-_{kl} +
\frac{1}{2} \pac{
\pa{\xi^+_{ik} + \sigma^2_i \delta_{ik}} \pa[2]{\xi^+_{jl} + \sigma^2_j \delta_{jl}} \cos4(\psi_{ij}-\psi_{kl})
+ \xi^-_{ik} \xi^-_{jl} \cos4(\psi_{ik}-\psi_{ij}+\psi_{jl}-\psi_{kl})
+ (k\leftrightarrow l)}
,
\\
\label{eq:Xi_+-}
\Xi^{+-}_{ijkl}
&= \xi^+_{ij} \xi^-_{kl} +
\frac{1}{2} \pac{
\pa{\xi^+_{ik} + \sigma_i^2 \delta_{ik}}  \xi^-_{jl} \cos4(\psi_{jl}-\psi_{kl})
+ \text{3 permutations ($i\leftrightarrow j$ or $k\leftrightarrow l$)}
}
.
\end{align}

\subsection{Sample averaging step}

The full covariance matrix reads
\begin{equation}
\Cov\pa{\hat{\xi}_{aIJ}^\sigma, \hat{\xi}_{bKL}^\eta}
=
\sev{
\frac{1}{P_{aIJ}P_{bKL}}
\sum_{(ij\in\pairs_{aIJ}} \sum_{(kl)\in\pairs_{bKL}}
\Xi^{\sigma\eta}_{ijkl}
}
-
\xi^\sigma_{\sev{12}_{aIJ}} \xi^\eta_{\sev{34}_{bKL}} \ .
\end{equation}
The application of the sample-averaging operator is then very similar to subsubsec.~\ref{subsubsec:mathematical_origin_sparsity_covariance}. I first divide the sum into terms that have zero, one, or two indices in common. Here the presence of potentially different redshift bins $I, J, K, L$ makes things somewhat more complicated, because for example $i=k$ is only possible if $I=K$, etc. Accounting for all possibilities,
\begin{multline}
\sev{
\frac{1}{P_{aIJ}P_{bKL}}
\sum_{(ij\in\pairs_{aIJ}} \sum_{(kl)\in\pairs_{bKL}}
\Xi^{\sigma\eta}_{ijkl}
}
=
F_0 \, \Xi^{\sigma\eta}_{\sev{12}_{aIJ} \sev{34}_{bKL}}
\\
+
F_{i=k} \, \Xi^{\sigma\eta}_{\sev{12}_{aIJ} \sev{14}_{bKL}}
+
F_{i=l} \, \Xi^{\sigma\eta}_{\sev{12}_{aIJ} \sev{14}_{bKL}}
+
F_{j=k} \, \Xi^{\sigma\eta}_{\sev{12}_{aIJ} \sev{14}_{bKL}}
+
F_{j=l} \, \Xi^{\sigma\eta}_{\sev{12}_{aIJ} \sev{14}_{bKL}}
\\
+ F_{(ij)=(kl)} \, \Xi^{\sigma\eta}_{\sev{12}_{aIJ} \sev{12}_{aKL}}
+ F_{(ij)=(lk)} \, \Xi^{\sigma\eta}_{\sev{12}_{aIJ} \sev{21}_{aKL}} \ ,
\end{multline}
where $F_{i=k}$ denotes the fraction of terms such that $i=k$, etc. The fraction of terms with no common indices is $F_0=1-F_{i=k}-F_{i=l}-\ldots$ The various fractions are estimated just like in subsubsec.~\ref{subsubsec:counting_pairs_of_pairs}. For instance, the number of terms with $i=k$ is
\begin{equation}
C_{i=k} = \delta _{IK} N_I \times \frac{N_J}{\Omega} \, \Omega_a \times \frac{N_L}{\Omega} \, \Omega_b
\end{equation}
while the total number of pairs of pairs in $\pairs_{aIJ}\times\pairs_{bKL}$ reads
\begin{equation}
P_{aIJ} P_{bKL} = \pa{N_I \times \frac{N_J}{\Omega} \, \Omega_a} \pa{N_K \times \frac{N_L}{\Omega} \, \Omega_b} ,
\end{equation}
so that
\begin{equation}
F_{i=k} = \frac{C_{i=k}}{P_{aIJ} P_{bKL}} = \frac{\delta_{IK}}{N_I} \ .
\end{equation}

Putting everything together, I get
\begin{align}
\Cov\pa{\hat{\xi}_{aIJ}^\sigma, \hat{\xi}_{bKL}^\eta}
&=
\underbrace{
\Xi^{\sigma\eta}_{\sev{12}_{aIJ} \sev{34}_{bKL}} - \xi^\sigma_{\sev{12}_{aIJ}} \xi^\eta_{\sev{34}_{bKL}}
}_{\CCov\pa{\hat{\xi}_{aIJ}^\sigma, \hat{\xi}_{bKL}^\eta}}
\\
&\quad +
\frac{\delta_{IK}}{N_I}
\pa{
\Xi^{\sigma\eta}_{\sev{12}_{aIJ} \sev{14}_{bIL}} - \Xi^{\sigma\eta}_{\sev{12}_{aIJ} \sev{34}_{bKL}}
}
+ \frac{\delta_{IL}}{N_I}
\pa{
\Xi^{\sigma\eta}_{\sev{12}_{aIJ} \sev{31}_{bKI}} - \Xi^{\sigma\eta}_{\sev{12}_{aIJ} \sev{34}_{bKL}}
}
\nonumber\\
&\quad
+ \frac{\delta_{JK}}{N_J}
\pa{
\Xi^{\sigma\eta}_{\sev{12}_{aIJ} \sev{24}_{bJL}} - \Xi^{\sigma\eta}_{\sev{12}_{aIJ} \sev{34}_{bKL}}
}
+ \frac{\delta_{JL}}{N_J}
\pa{
\Xi^{\sigma\eta}_{\sev{12}_{aIJ} \sev{32}_{bIL}} - \Xi^{\sigma\eta}_{\sev{12}_{aIJ} \sev{34}_{bKL}}
}
\nonumber\\
&\quad
\underbrace{
+
\frac{\delta_{ab} \delta_{IK}\delta_{JL}}{P_{aIJ}}
\pa{ \Xi^{\sigma\eta}_{\sev{12}_{aIJ} \sev{12}_{aIJ}}  - \Xi^{\sigma\eta}_{\sev{12}_{aIJ} \sev{34}_{bKL}} }
+
\frac{\delta_{ab} \delta_{IL}\delta_{JK}}{P_{aIJ}}
\pa{ \Xi^{\sigma\eta}_{\sev{12}_{aIJ} \sev{21}_{aJI}}  - \Xi^{\sigma\eta}_{\sev{12}_{aIJ} \sev{34}_{bKL}} }
}_{(\SCov+\NCov)\pa{\hat{\xi}_{aIJ}^\sigma, \hat{\xi}_{bKL}^\eta}}
,
\nonumber
\end{align}
where the first line, which is independent of sample size, is cosmic covariance, while the other terms encode both sparsity and noise covariances. The latter two may be conventionally distinguished by defining $\SCov$ as the terms of $\SCov+\NCov$ that do not depend on $\sigma^2$.

\subsection{Explicit expressions for all covariance terms}

Explicit expressions for the covariance matrix elements are obtained by applying the various sample averaging schemes on the indices of $\Xi^{\sigma\eta}_{ijkl}$. For example, $\Xi^{\sigma\eta}_{\sev{12}_{aIJ} \sev{14}_{bIL}}$ means that $\Xi^{\sigma\eta}_{1214}$ is averaged over $z_1\in\redshiftbin_I$ and $\uvect{u}_1\in\footprint$; $z_2\in\redshiftbin_J$ and $\uvect{u}_2\in\annulus_a(\uvect{u}_1)$; and $z_4\in\redshiftbin_L$ and $\uvect{u}_4\in\annulus_b(\uvect{u}_1)$, that is
\begin{multline}
\Xi^{\sigma\eta}_{\sev{12}_{aIJ} \sev{14}_{bIL}}
=
\int_0^\infty \dd z_1 \; p_I(z_1)
\int_\footprint \frac{\dd^2\uvect{u}_1}{\Omega}
\int_0^\infty \dd z_2 \; p_J(z_2)
\int_{\annulus_a(\uvect{u}_1)} \frac{\dd^2\uvect{u}_2}{\Omega_a(\uvect{u}_1)}
\int_0^\infty \dd z_4 \; p_L(z_4)
\int_{\annulus_b(\uvect{u}_1)} \frac{\dd^2\uvect{u}_4}{\Omega_a(\uvect{u}_1)}
\\
\times
\Xi^{\sigma\sigma}(\uvect{u}_1, z_1 ; \uvect{u}_2, z_2; \uvect{u}_1, z_1; \uvect{u}_4, z_4) \ .
\end{multline}
Working in the pseudo-full sky approximation, all positions of $\uvect{u}_1$ are equivalent thanks to statistical isotropy; I may simply place it at the North pole, $\uvect{u}_1=\uvect{z}$, and omit the $\uvect{u}_1$ dependence in $\annulus_a, \annulus_b$. This equally applies to $\uvect{u}_2, \uvect{u}_3, \uvect{u}_4$, respectively,  when $2, 3, 4$ is the common index.

\subsubsection{Cosmic covariance}

The explicit expression of the cosmic covariance of cosmic shear, in the pseudo-full-sky and flat-sky approximations, is
\begin{align}
\label{eq:cosmic_covariance_shear_++}
\CCov\pa{\hat{\xi}_{aIJ}^+, \hat{\xi}_{bKL}^+}
&=
\frac{1}{2}
\int_\footprint \frac{\dd^2\uvect{u}_3}{\Omega}
\int_{\annulus_a} \frac{\dd^2\uvect{u}_2}{\Omega_a}
\int_{\annulus_b} \frac{\dd^2\uvect{u}_4}{\Omega_b} \;
\Big[
\xi^+_{IK}(\theta_3)
\xi^+_{JL}(\theta_{24})
+
\xi^-_{IK}(\theta_3)
\xi^-_{JL}(\theta_{24})
\cos4(\psi_{24}-\ph_3)
\Big]
+ (K\leftrightarrow L),
\\
\label{eq:cosmic_covariance_shear_--}
\CCov\pa{\hat{\xi}_{aIJ}^-, \hat{\xi}_{bKL}^-}
&=
\frac{1}{2}
\int_\footprint \frac{\dd^2\uvect{u}_3}{\Omega}
\int_{\annulus_a} \frac{\dd^2\uvect{u}_2}{\Omega_a}
\int_{\annulus_b} \frac{\dd^2\uvect{u}_4}{\Omega_b} \;
\Big[
\xi^+_{IK}(\theta_3)
\xi^+_{JL}(\theta_{24})
\cos4(\ph_2 - \ph_4)
\\ \nonumber
&\quad
+
\xi^-_{IK}(\theta_3)
\xi^-_{JL}(\theta_{24})
\cos4(\psi_{24} + \ph_3 - \ph_1 - \ph_2)
\Big]
+ (K\leftrightarrow L),
\\
\label{eq:cosmic_covariance_shear_+-}
\CCov\pa{\hat{\xi}_{aIJ}^+, \hat{\xi}_{bKL}^-}
&=
\frac{1}{2}
\int_\footprint \frac{\dd^2\uvect{u}_3}{\Omega}
\int_{\annulus_a} \frac{\dd^2\uvect{u}_2}{\Omega_a}
\int_{\annulus_b} \frac{\dd^2\uvect{u}_4}{\Omega_b} \;
\xi^-_{IK}(\theta_3)
\xi^+_{JL}(\theta_{24})
\cos4(\ph_3 - \ph_2)
\\ \nonumber
&\quad
+ \text{3 permutations ($I\leftrightarrow J$ or $K\leftrightarrow L$)} ,
\end{align}
with the shorthand notation
\begin{equation}
\xi^\pm_{IJ}(\theta)
\define
\int_0^\infty\dd z_1 \; p_I(z_1) \int_0^\infty \dd z_2 \; p_J(z_2) \, \xi^\pm(\theta; z_1, z_2) \ .
\end{equation}
\Cref{eq:cosmic_covariance_shear_++,eq:cosmic_covariance_shear_--,eq:cosmic_covariance_shear_+-} match eqs.~(34), (36) and (38) of \cite{2002A&A...396....1S} up to changes of variables in the integrals, and further applications of the flat-sky approximation to express the angles $\theta_{24}, \psi_{24}$.

\subsubsection{Noise and sparsity covariances}

The explicit expression of the sparsity plus noise covariance of cosmic shear, in the pseudo-full-sky and flat sky approximations, is
\begin{align}
\label{eq:sparsity_noise_covariance_shear_++_complete}
&(\SCov+\NCov)\pa{\hat{\xi}_{aIJ}^+, \hat{\xi}_{bKL}^+}
\\
&=
\frac{\delta_{IK}}{N_I}
\frac{1}{2}
\int_{\annulus_a} \frac{\dd^2\uvect{u}_2}{\Omega_a}
\int_{\annulus_b} \frac{\dd^2\uvect{u}_4}{\Omega_b}
\paac{
\pac{\xi^+_I(0) + \sigma_I^2} \xi^+_{JL}(\theta_{24})
+
\negligible{
3(\xi^+ \xi^+)_{IJL}(\theta_2, \theta_4)
}
}
\nonumber\\ &\qquad
-
\negligible{
\frac{\delta_{IK}}{N_I}
\pac{
\int_{\annulus_a} \frac{\dd^2\uvect{u}_2}{\Omega_a} \; \xi^+_{IJ}(\theta_2) 
\int_{\annulus_b} \frac{\dd^2\uvect{u}_4}{\Omega_b} \; \xi^+_{IL}(\theta_4)
+
\CCov\pa{\hat{\xi}_{aIJ}^+, \hat{\xi}_{bKL}^+}
}
}
+ \text{3 permutations ($I\leftrightarrow J$ or $K\leftrightarrow L$)}
\nonumber\\ & \quad
+
\frac{\delta_{ab}\delta_{IK}\delta_{JL} }{P_{aIJ}}
\frac{1}{2}
\paac{
\pac{\xi^+_I(0) + \sigma_I^2} \pac{\xi^+_J(0) + \sigma_J^2}
+
\negligible{
\int_{\annulus_a} \frac{\dd^2 \uvect{u}_2}{\Omega_a}
\pac{3(\xi^+\xi^+)_{IJ}(\theta_2) + (\xi^-\xi^-)_{IJ}(\theta_2)}}
}
\nonumber\\ &\qquad
-
\negligible{
\frac{\delta_{ab}\delta_{IK}\delta_{JL} }{P_{aIJ}}
\paac{
\pac{
\int_{\annulus_a} \frac{\dd^2\uvect{u}_2}{\Omega_a} \; \xi^+_{IJ}(\theta_2) 
}^2
+
\CCov\pa{\hat{\xi}_{aIJ}^+, \hat{\xi}_{bKL}^+}
}
}
+ \text{1 permutation ($K\leftrightarrow L$)},
\nonumber
\end{align}
\begin{align}
\label{eq:sparsity_noise_covariance_shear_--_complete}
&(\SCov+\NCov)\pa{\hat{\xi}_{aIJ}^-, \hat{\xi}_{bKL}^-}
\\
&=
\frac{\delta_{IK}}{N_I}
\frac{1}{2}
\int_{\annulus_a} \frac{\dd^2\uvect{u}_2}{\Omega_a}
\int_{\annulus_b} \frac{\dd^2\uvect{u}_4}{\Omega_b}
\paac{
\pac{\xi^+_I(0) + \sigma_I^2} \xi^+_{JL}(\theta_{24}) \cos4(\ph_2 - \ph_4)
+ 
\negligible{
3(\xi^- \xi^-)_{IJL}(\theta_2, \theta_4)
}
}
\nonumber\\ &\qquad
-
\negligible{
\frac{\delta_{IK}}{N_I}
\pac{
\int_{\annulus_a} \frac{\dd^2\uvect{u}_2}{\Omega_a} \; \xi^-_{IJ}(\theta_2) 
\int_{\annulus_b} \frac{\dd^2\uvect{u}_4}{\Omega_b} \; \xi^-_{IL}(\theta_4)
+
\CCov\pa{\hat{\xi}_{aIJ}^-, \hat{\xi}_{bKL}^-}
}
}
+ \text{3 permutations ($I\leftrightarrow J$ or $K\leftrightarrow L$)}
\nonumber\\ & \quad
+
\frac{\delta_{ab}\delta_{IK}\delta_{JL} }{P_{aIJ}}
\frac{1}{2}
\paac{
\pac{\xi^+_I(0) + \sigma_I^2} \pac{\xi^+_J(0) + \sigma_J^2}
+
\int_{\annulus_a} \frac{\dd^2 \uvect{u}_2}{\Omega_a}
\pac{(\xi^+\xi^+)_{IJ}(\theta_2) + 3(\xi^-\xi^-)_{IJ}(\theta_2)}
}
\nonumber\\ &\qquad
-
\negligible{
\frac{\delta_{ab}\delta_{IK}\delta_{JL} }{P_{aIJ}}
\paac{
\pac{
\int_{\annulus_a} \frac{\dd^2\uvect{u}_2}{\Omega_a} \; \xi^-_{IJ}(\theta_2) 
}^2
+
\CCov\pa{\hat{\xi}_{aIJ}^-, \hat{\xi}_{bKL}^-}
}
}
+ \text{1 permutation ($K\leftrightarrow L$)},
\nonumber
\end{align}
\begin{align}
\label{eq:sparsity_noise_covariance_shear_+-_complete}
&(\SCov+\NCov)\pa{\hat{\xi}_{aIJ}^+ , \hat{\xi}_{bKL}^-}
\\
&=
\frac{\delta_{IK}}{N_I}
\frac{1}{2}
\int_{\annulus_a} \frac{\dd^2\uvect{u}_2}{\Omega_a}
\int_{\annulus_b} \frac{\dd^2\uvect{u}_4}{\Omega_b}
\paac{
\pac{\xi^+_I(0) + \sigma_I^2} \xi^-_{JL}(\theta_{24}) \cos4\psi_{24}
+
\negligible{
3(\xi^+ \xi^-)_{IJL}(\theta_2, \theta_4)
}
}
\nonumber\\ &\qquad
-
\negligible{
\frac{\delta_{IK}}{N_I}
\pac{
\int_{\annulus_a} \frac{\dd^2\uvect{u}_2}{\Omega_a} \; \xi^+_{IJ}(\theta_2) 
\int_{\annulus_b} \frac{\dd^2\uvect{u}_4}{\Omega_b} \; \xi^-_{IL}(\theta_4)
+
\CCov\pa{\hat{\xi}_{aIJ}^+, \hat{\xi}_{bKL}^-}
}
}
+ \text{3 permutations ($I\leftrightarrow J$ or $K\leftrightarrow L$)}
\nonumber\\ & \quad
+
\negligible{
\frac{\delta_{ab}\delta_{IK}\delta_{JL} }{P_{aIJ}}
\int_{\annulus_a} \frac{\dd^2\uvect{u}_2}{\Omega_a} \;
2(\xi^+\xi^-)_{IJ}(\theta_2)
}
\nonumber\\ &\qquad
-
\negligible{
\frac{\delta_{ab}\delta_{IK}\delta_{JL} }{P_{aIJ}}
\pac{
\int_{\annulus_a} \frac{\dd^2\uvect{u}_2}{\Omega_a} \; \xi^+_{IJ}(\theta_2)
\int_{\annulus_a} \frac{\dd^2\uvect{u}_4}{\Omega_a} \; \xi^-_{IJ}(\theta_4)
+
\CCov\pa{\hat{\xi}_{aIJ}^+, \hat{\xi}_{bKL}^-}
}
}
+ \text{1 permutation ($K\leftrightarrow L$)} ,
\nonumber
\end{align}
with the shorthand notation
\begin{align}
(\xi^\sigma\xi^\eta)_{IJK}(\theta, \theta')
&=
\int_0^\infty\dd z_1 \; p_I(z_1) \int_0^\infty \dd z_2 \; p_J(z_2) \int_0^\infty\dd z_3 \; p_K(z_3) \,
\xi^\sigma(\theta; z_1, z_2) \xi^\eta(\theta'; z_1, z_3) \ ,
\\
(\xi^\sigma\xi^\eta)_{IJ}(\theta)
&=
\int_0^\infty\dd z_1 \; p_I(z_1) \int_0^\infty \dd z_2 \; p_J(z_2)
\xi^\sigma(\theta; z_1, z_2) \xi^\eta(\theta; z_1, z_2) \ .
\end{align}
In \cref{eq:sparsity_noise_covariance_shear_++_complete,eq:sparsity_noise_covariance_shear_--_complete,eq:sparsity_noise_covariance_shear_+-_complete}, all the terms written in \negligible{grey} are of the form
\begin{equation}
\frac{1}{\text{large number}} \pac{\int \frac{\dd^2\uvect{u}}{\Omega_a} \; \xi^\pm}^2 \sim \frac{\CCov}{\text{large number}} \ .
\end{equation}
As such, they are negligible compared to cosmic covariance in practice. The terms that survive feature the mean square of intrinsic ellipticities, $\sigma^2$, and $\xi^+(0)\gg \int\frac{\dd^2\uvect{u}}{\Omega_a} \; \xi^\pm$; they dominate noise and sparsity covariances for cosmic shear.

\end{widetext}

\bibliographystyle{aasjournal}
\bibliography{bibliography.bib}

\end{document}